\documentclass[a4paper,fleqn]{cas-dc}
\usepackage[numbers]{natbib}
\usepackage{tabularx} 
\def\tsc#1{\csdef{#1}{\textsc{\lowercase{#1}}\xspace}}
\tsc{WGM}
\tsc{QE}
\tsc{EP}
\tsc{PMS}
\tsc{BEC}
\tsc{DE}

\begin{document}
\let\WriteBookmarks\relax
\def\floatpagepagefraction{1}
\def\textpagefraction{.001}

\shorttitle{First-principles study of structural, electronic and optical properties of non-toxic RbBaX$_3$}

\shortauthors{Pranti Saha et~al.}

\title [mode = title]{First-principles study of structural, electronic and optical properties of non-toxic RbBaX$_3$ (X = F, Cl, Br, I) perovskites under hydrostatic pressure}                      



%
\author[1]{Pranti Saha}[type=editor,
                        auid=000,bioid=1,
                        orcid=0009-0000-0913-3078]

\cormark[1]


\ead{saha.pranti2000@gmail.com}


\credit{Conceptualization, methodology, software simulation, formal analysis, writing - original draft, reviewing and editing}

\affiliation[1]{organization={Department of Materials Science and Engineering},
    addressline={Khulna University of Engineering \& Technology (KUET)}, 
    city={Khulna},
    citysep={}, 
    postcode={9203}, 
    country={Bangladesh}}

\author[2]{In Jun Park}[]

\affiliation[2]{organization={Department of Physics, Indiana University Indianapolis},
    city={Indianapolis},
    citysep={}, 
    postcode={46202}, 
    state={Indiana},
    country={USA}}

\credit{Formal analysis, data curation, reviewing and editing}

\author[3]{Protik Das}[]

\credit{Methodology, Formal analysis, reviewing and editing}

\affiliation[3]{organization={KLA AI Infrastructure Design group},
    city={Milpitas},
    citysep={}, 
    postcode={95035}, 
    state={California},
    country={USA}}

\author[4]{Fariborz Kargar}[orcid=0000-0003-3361-7894]

\affiliation[4]{organization={Materials Research and Education Center, Department of Mechanical Engineering, Auburn University},
    city={Auburn},
    citysep={}, 
    postcode={36849}, 
    state={Alabama},
    country={USA}}

\credit{Conceptualization, supervision, formal analysis, validation, reviewing and editing}

\cortext[cor1]{Corresponding author}



\begin{abstract}
We have investigated the structural, mechanical, electronic and optical properties of Rb-based cubic perovskite RbBaX$_3$ (X = F, Cl, Br, I) under hydrostatic pressure, using first-principle density functional theory (DFT). 
All RbBaX$_3$ perovskites exhibit thermodynamic and mechanical stability at ambient pressure.
RbBaF$_3$ remains structurally stable across all examined pressures, while RbBaCl$_3$, RbBaBr$_3$, and RbBaI$_3$ maintain mechanical stability up to 60, 60, and 40 GPa, respectively.
These materials are ductile even at elevated pressure.
RbBaF$_3$ has a direct bandgap of 4.80 eV while other compositions exhibit indirect band gaps of 4.37, 3.73, and 3.24 eV with halide atoms of Cl, Br, and I, respectively.
Under elevated hydrostatic pressure, only RbBaCl$_3$ and RbBaI$_3$ exhibit an indirect-to-direct band transition while others preserve their nature of band gap. 
Our results show that spin-orbit coupling significantly affects only the valance bands of larger-sized halides (Cl, Br, I).
With hybrid functional (HSE) correction, the band gaps of these four materials increase to 6.7, 5.6, 4.8 and 4.4 eV, respectively, but the nature of direct/indirect band transition remains unchanged. 
Orbital-decomposed partial density of states calculation reveals that the halogen \textit{p}-orbitals dominate the valence band near the Fermi level, while Rb 5\textit{s}-orbital affects the conduction band minima the most.
Investigation of the optical properties reveals wide-band absorption, low electron loss, moderate reflectivity and lower refractive index in the UV to deep-UV range.
The strength and range of absorption increases significantly with hydrostatic pressure, suggesting that RbBaX$_3$ perovskites are promising candidates for tunable UV-absorbing optoelectronic devices.  
\end{abstract}




\maketitle

\section{Introduction}
Metal halide perovskites have attracted significant
attention since the 1950s \cite{moller1958crystal}, owing to their exceptional semiconducting and 
optoelectronic properties, including broad absorption spectra \cite{jiang2020ultra, liu2020templated}, enhanced optical absorption \cite{2020_nanolett_optical},
tunable band gaps \cite{peedikakkandy2018lead, haque2018origin, chen2021advances}, long charge diffusion length \cite{yin2020toward, jeong2020stable}, high defect tolerance \cite{2017_defect_tol}, high charge carrier mobility \cite{2017_mobility, motta2015charge}, and low carrier effective mass \cite{2013_effective_Mass, miyata2015direct}. 
Such properties are promising for a wide range of applications such as solar cells \cite{CsPbCl3_pressure}, thermoelectric devices \cite{he2014perovskites, mettan2015tuning}, light emitting diodes \cite{CsPbX3_exp, stranks2015metal}, and solid-state memory \cite{memory}.
For instance, the photoelectric conversion efficiency of solar cells based on organic-inorganic hybrid halide perovskites, particularly methylammonium lead iodide (MAPbI$_3$), has recently improved from 3.8\% to 22\%, opening new avenues for low-cost solar energy devices \cite{jeong2022large, lin2022all, xiao2020all}. 
However, concerns remain regarding the lead-based hybrid perovskites, including the thermodynamic breakdown of the organic components and more importantly, the toxic implications of the Pb ion leaking to the environments \cite{ren2022potential}. 
In contrast, the inorganic perovskites offer greater stability at high temperatures and humidity, largely due to the suppression of the halide segregation \cite{wang2019review, lee2015formamidinium, jena2019halide, jin2021can}.
As a result, inorganic halide perovskites have emerged as eco-friendly alternatives for stable optoelectronic applications \cite{roknuzzaman2017towards}.

Metal halide perovskites are represented by the chemical formula of ABX$_3$, where A is a monovalent cation, B is a divalent cation, and X is a halogen atom.
Structurally, A atom occupies the decahedral sites, with B atoms octahedrally coordinated to the X atoms \cite{das2022density}. 
Numerous combinations of ABX$_3$ compositions are possible, with A and B atoms selected from alkali metals (Li, Na, K, Rb, Cs), alkaline-earth metals (Be, Mg, Ca, Sr, Ba), or other elements (Ag, Zn, Cd, Hg, Pb) \cite{Nishimatsu_2002}.
RbBaX$_3$ is one such member of the halide perovskite family that is devoid of Pb toxicity and is the focus of the current study.
Many halide perovskites exhibit temperature-driven phase transitions. 
For example, CsSnI$_3$ perovskite undergoes several phase transitions from the cubic $\alpha$ phase to tetragonal $\beta$, to orthorhombic $\gamma$ and non-perovskites Y-phase, as the temperature decreases \cite{CsSnX3_CsPbX3}.
Given that Rb is in the same alkali metal group as Cs, Rb-based perovskites are expected to exhibit several phases as Cs-based perovskites, and thus Rb is another promising candidate for the A$^{+}$ cation replacement \cite{ULLAH2022106977}. 
In this paper, however, we focus our theoretical analysis on the high-temperature cubic phase of Rb-based perovskites.

The formability and the stability of cubic RbBaX$_3$ perovskite 
can be assessed by the Goldschmidt tolerance factor \cite{goldschmidt1926akad}, $t=(r_A+r_X)/\sqrt{2}(r_B+r_X)$, 
where $r_A$, $r_B$, and $r_X$ are the ionic radii of cation A, cation B, and the halogen X.
For a stable perovskite, $t$ should generally fall between 0.8 and 1.1, 
though this range has later been revised to $0.75\leq t \leq 1.0$ \cite{paul2023synthesis, tuning_tolerance}.
Considering Shannon’s effective ionic radii \cite{Nishimatsu_2002} to calculate the tolerance factor, one would obtain \textit{t} values of 0.81, 0.79, 0.78, and 0.78, for RbBaF$_3$, RbBaCl$_3$, RbBaBr$_3$, RbBaI$_3$, respectively, 
indicating that these materials are near the stability criteria range \cite{2024_mini_review}.
In addition to the Goldschmidt factor, the octahedral factor (OF), $\mu = r_B/r_X$, provides another stability criterion by considering the tilting between the Ba and X sites \cite{2024_mini_review}. 
For stable perovskite, the octahedral factor $\mu$ should be between 0.44 and 0.9.
For RbBaX$_3$, the values of $\mu$ are 1.0, 0.75, 0.69, and 0.62 for F, Cl, Br, and I, respectively.
%
An alternative and more precise way to estimate the stability of the cubic phase of these perovskites is to calculate the formation energy. 
A negative formation energy indicates that the phase is thermodynamically stable and energetically more favorable \cite{pingak2023lead}.

Several experimental and theoretical groups have investigated the RbBX$_3$ perovskites to better understand their structural, elastic, electronic, optical, transport and thermodynamic properties. 
Early experimental studies successfully demonstrated methods for the synthesis of stable RbPbCl$_3$ and RbSnF$_3$ \cite{paul2023synthesis,2014_RbSnF3}.
Theoretical calculations have reported the band gap of several RbBF$_3$ perovskites (B = Be, Mg, Ca, Sr, Ba, Hg, Pb) using the local density approximation (LDA) theory and predicted that most of these structures are indirect semiconductors.
These calculations predicted that
RbBaF$_3$ has an LDA band gap of 5.60 eV \cite{Nishimatsu_2002}. 
Note that LDA generally suffers from ‘over-binding’ the bond length which underestimates the band gap. 
A subsequent \textit{ab initio} study \cite{RbBaF3_GW} showed that the LDA band gap of RbBaF$_3$ is 4.39 eV, which increased to 8.05 eV when the more accurate though computationally costly quasi-particle GW correction was employed.
Their first principle calculation indicates that RbBaF$_3$ should be a direct band gap material, contrary to  earlier findings \cite{Nishimatsu_2002}.
Another first-principle calculation based on generalized gradient approximation (GGA) reported that cubic RbBaF$_3$ has an indirect band gap of 5.6 eV, which increases to 9.2 eV with the modified Becke-Johnson exchange potential approximation \cite{2017_RbMF3}.
It is clear that there is still some ambiguity in the research community about whether the band transition in RbBaF$_3$ perovskite is direct or indirect.  

Despite the existing theoretical and experimental literature on Rb-based perovskites, many of their properties remain poorly understood. 
The mechanical stability, ductility, and elastic moduli have largely been unexplored through first-principle calculations. 
Although a few ab initio simulations of RbBaF$_3$ under ambient pressure conditions exist in the literature, no previous studies have investigated the band profiles of these structures with larger halogen atoms of Cl, Br, and I using density functional theory, either at ambient conditions or under hydrostatic pressure. 
Additionally, the complex dielectric response and the optical properties of RbBaCl$_3$, RbBaBr$_3$, and RbBaI$_3$ are notably absent in the literature. 
Addressing these gaps is crucial for advancing the design and application of non-toxic Rb-based perovskites in various technological fields.

In this paper, we investigate the  structural, elastic, mechanical,
electronic, and optical properties of RbBaX$_3$ (X = F, Cl, Br, I) under hydrostatic pressure.
Particularly, we address the impact of hydrostatic pressure on the nature of band transition in RbBaX$_3$ structures, focusing on the direct vs. indirect band gap controversy in RbBaF$_3$.
The band profiles are evaluated using more accurate hybrid functional (HSE) formalism to provide a more realistic band gap for these perovskites. 
We also consider the spin-orbit coupling (SOC) effects which were overlooked in the earlier studies. 

The paper is organized as follows:
we first brieﬂy summarize the computational details of density functional theory simulation employed in this work. 
Then, we discuss the relaxed atomic geometry, thermodynamic stability and mechanical properties of the four halide perovskites under hydrostatic pressure, commenting on their stability based on the Born stability criteria.
Next, we present the band structure and the electronic density of states as one moves from fluoride to iodide-based perovskites. 
We quantify the effect of SOC, hybrid functional correction and external pressure on the nature of band-to-band transitions and the resultant band gap energy.
Finally, we assess the complex dielectric function and optical response of this Rb class of perovskites, discussing their potential as candidates for optoelectronic applications.

\section{Computational Methodology}
To calculate the structural, electronic and optical properties of the
inorganic perovskite RbBaX$_3$, we perform the density functional theory calculations, as implemented in the Quantum Espresso code \cite{Giannozzi_2009, Giannozzi_2017}.
The Optimized Norm-Conserving Vanderbilt (ONCV) SG15 pseudopotentials set \cite{ONCV1,ONCV2} has been used, at the level of the Perdew-Burke-Ernzerhof (PBE) exchange-correlation theory. 
However, the structure relaxation and the band profile remain almost the same when we use the projector-augmented-wave (PAW) pseudo-potential from PS Library \cite{dal2013ab}, indicating that the results are invariant to the choice of the pseudopotentials.
The structure relaxation was carried out until the force on each atom became lower than 0.001 Rydberg/Bohr.
A convergence threshold of 10$^{-10}$ atomic unit (Rydberg) is chosen for the self-consistency simulations.
The formation energy $\Delta E_f$ for RbBaX$_3$ is calculated using the formula \cite{pingak2023lead},
\begin{equation}
\Delta E_f = \frac{E_{\mathrm{RbBaX_3}} - E_{\mathrm{Rb}} - E_{\mathrm{Ba}} - 3 E_{\mathrm{X}}}{N},    
\end{equation}
where E$_i$ indicates the energy of the chemical entity $i$, and N is total number of atoms in the unit cell.

The kinetic energy and the charge density cut-off parameters are set to 60 Rydberg and 480 Rydberg, respectively.  
The k-points grid for the self-consistent field (SCF) and non-self-consistent field (NSCF) computations were set to 8 × 8 × 8 and 14 × 14 × 14, respectively.
The energy and charge density cut-offs, as well as the k-points have been checked for convergence for all the materials.
For the Heyd-Scuseria-Ernzerhof (HSE) hybrid functional correction, 
we applied HSE06 formalism with the HF mixing fraction of 0.25.

The spin-orbit coupling in the band structure calculation is considered through the non-collinear implementation. 
This will improve the accuracy of the band structure, specifically for the heavier atoms like Br and I, for which the spin-orbit coupling is an important effect \cite{das2022density,OBADA2024108372}.
To include the SOC effect, we use the fully-relativistic version of the pseudopotentials.
However, since it is computationally expensive, we do not incorporate the SOC and HSE correction in extracting the mechanical and optical properties.

The complex dielectric function, $\epsilon = \epsilon_1 + i\epsilon_2$, where $\epsilon_1$ and $\epsilon_2$ are the real and imaginary components of the response, is calculated 
using the time-dependent first-order perturbation theory \cite{PhysRevB.67.165332}. 
The real function ($\epsilon_1$) is calculated from the Kramers–Kronig transformation relations \cite{peiponen2004kramers}, whereas 
the imaginary function ($\epsilon_2$) is obtained from the momentum matrix calculations \cite{xu1990determination}.
Convergence of dielectric function with respect to the k-grid and number of bands is ensured.
%
The absorption coefficient $\alpha(\omega)$, the electron loss function $L(\omega)$, refractive index $n(\omega)$, and reflectivity $R(\omega)$ are calculated from the simulated dielectric response \cite{LASHGARI201676, epsilon_formalism2}, 
\begin{align}
    \alpha(\omega) &= \frac{\sqrt{2}\omega}{c} \Big[\sqrt{ \epsilon_1(\omega)^2 + \epsilon_2(\omega)^2} - \epsilon_1(\omega)\Big]^{1/2}, \\ 
    L(\omega) &= \frac{ \epsilon_2(\omega) }{ \epsilon_1(\omega)^2 + \epsilon_2(\omega)^2 }, \\ 
    n(\omega) &= \frac{1}{\sqrt{2}}\Bigg[\sqrt{ \epsilon_1(\omega)^2 + \epsilon_2(\omega)^2} + \epsilon_1(\omega)\Bigg]^{1/2}, \\
    R(\omega) &= \frac{(n-1)^2+k^2}{(n+1)^2+k^2},\,\, k = \epsilon_2/2n(\omega)
\end{align}
 \\

\begin{figure}[t]
\centering 
\includegraphics[width=2.0in]{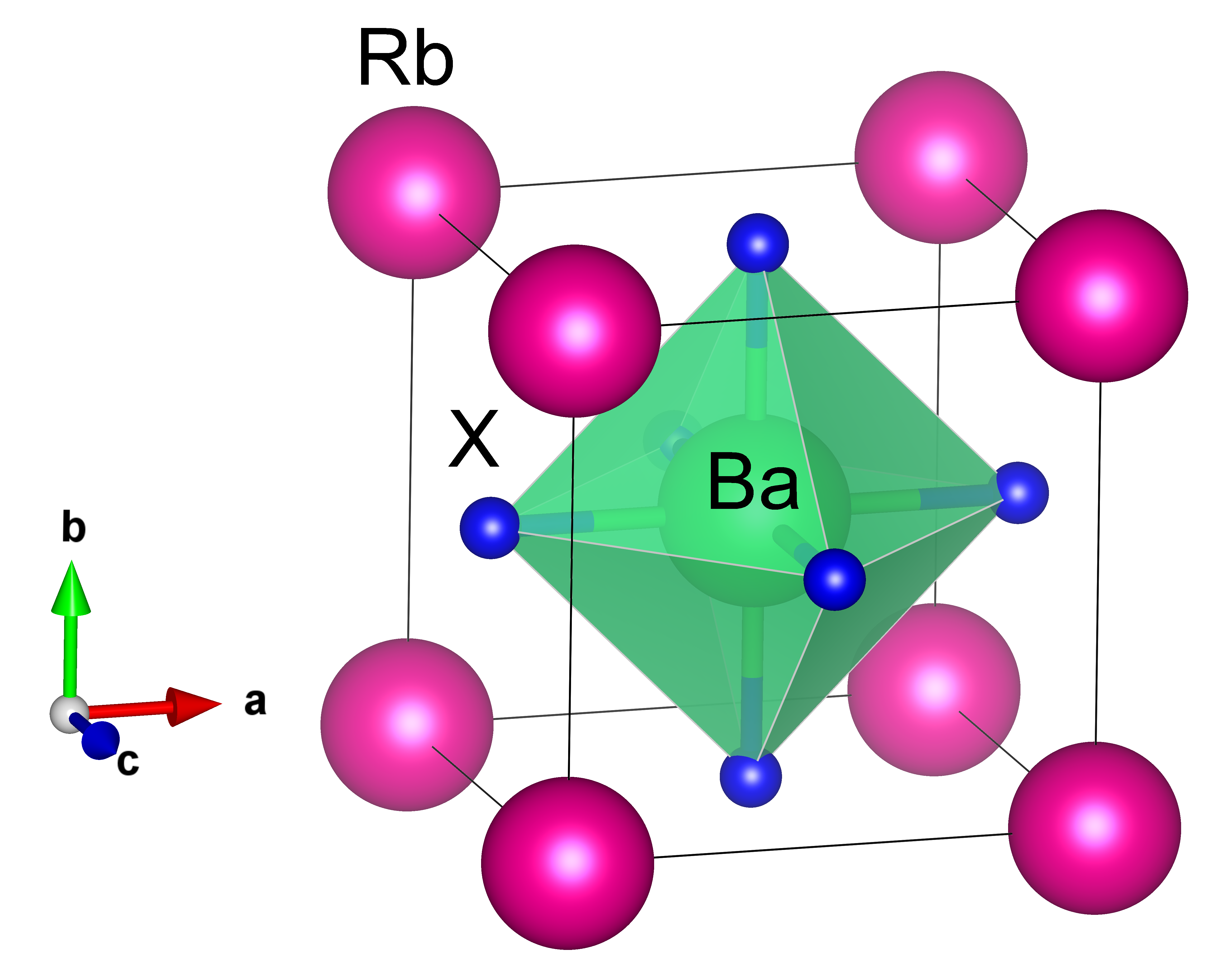}
\caption{Optimized structure of cubic perovskite RbBaX$_3$, where X is either F, Cl, Br or I. 
}
\label{fig:atomic_model}
\end{figure}

\begin{figure}[b]
\centering 
\includegraphics[height=3.8in]{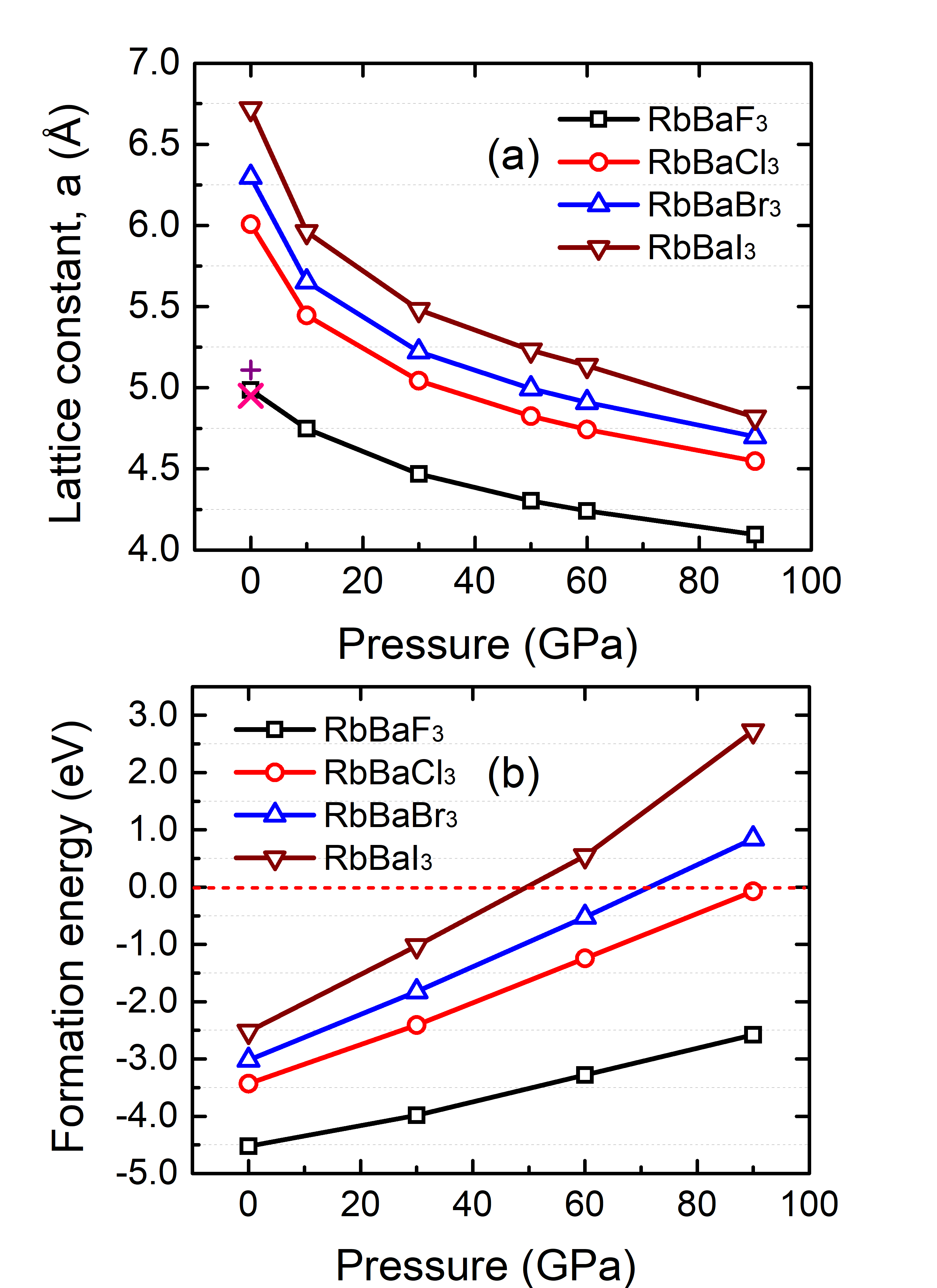}
\caption{(a) DFT-optimized lattice constant (\AA) of RbBaX$_3$ for different hydrostatic pressure up to 90 GPa. The $+$ and $\times$ symbols indicate lattice constants of RbBaF$_3$ from earlier DFT reports. (b) Formation energy $\Delta E_f$ of RbBaX$_3$ for different pressure conditions. The dash line indicates the formation energy line below which the structures are thermodynamically stable.}
\label{fig:lattice_vs_pressure}
\end{figure}

\section{Results and discussions}
\subsection{Structural properties}
The RbBaX$_3$ perovskite has a crystal structure with Rb cation placed within the BaX$_3$ octahedra.
Changing the halide atom at the corner-sharing BX$_3$ unit will change the structural and chemical properties of the perovskite \cite{Halide_perovskites_perspective, B403482A}.
The perovskites typically exhibit cubic phase at high temperatures, while other lower symmetry phases such as tetragonal, orthorhombic, monoclinic and/or rhombohedral, are observed at lower temperature ranges \cite{yu2019advances}.
For the RbBaX$_3$, we consider the primitive cubic form (Bravais lattice cP) which has Pm$\Bar{3}$m space group symmetry (number 221).
Figure \ref{fig:atomic_model} shows the relaxed structure of RbBaX$_3$.
The halide (X) atoms are positioned at Wyckoff positions 3c, namely, (0, 0.5, 0.5), (0.5, 0, 0.5), and (0.5, 0.5, 0). 
On the other hand, the Rb and Ba atoms are positioned at 1a (0, 0, 0) and 1b (0.5, 0.5, 0.5), respectively.

Figure \ref{fig:lattice_vs_pressure} (a) shows the evolution of the lattice constant as a function of hydrostatic pressure for various RbBaX$_3$ structures. 
At zero pressure, the computed lattice constants for RbBaF$_3$, RbBaCl$_3$, RbBaBr$_3$ and RbBaI$_3$ perovskites are 4.99 \AA, 6.00 \AA, 6.29 \AA\, and 6.72 \AA, respectively.
Replacing the X site with a heavier and larger species of halogen (F $\rightarrow$ Cl $\rightarrow$ Br $\rightarrow$ I) increases the lattice parameter and inter-atomic bond distance by 20\% $\sim$ 35\%.
This trend with changing halogen size and charge is consistent with other Cs-based halide perovskites \cite{roknuzzaman2017towards}. 
A prior \textit{ab initio} study with GGA formalism has reported the relaxed lattice constant of 5.11\AA\, for RbBaF$_3$ \cite{RbBaF3}, which is close to our optimized lattice parameter. 
A more accurate study with LDA+GW formalism calculated the lattice constant to be 4.95 \AA\, 
 \cite{RbBaF3_GW}, which is even closer to our calculated lattice constant of 4.99 \AA. 
For other halides (Br, Cl, I), there are no prior DFT simulation or experimental reports in the literature to compare with.
However, there exists a neural network based model which predicted the lattice constant of RbBaBr$_3$ to be 6.21 \AA\, \cite{2021_ANN}. 
This is close to our full DFT simulated lattice parameters of 6.29 \AA, essentially providing the validity of existing machine learning model. 
Table \ref{table:lattice_constant} summarizes the lattice constants of all the RbBaX$_3$ perovskites for hydrostatic pressure up to 90 GPa.

To understand the thermodynamic stability of these materials, we have calculated the formation energy, $\Delta E_f$.
Figure \ref{fig:lattice_vs_pressure} (b) presents the formation energy of RbBaX$_3$ as a function of the hydrostatic pressure.
As evident from the figure, all the materials are thermodynamically 
stable at ambient pressure condition. 
However, as the pressure  increases, $\Delta E_f$ becomes positive for all perovskites except RbBaF$_3$, indicating that the formation of those perovskites with $\Delta E_f$ > 0 is no longer favorable. 
RbBaF$_3$ is the only structure that remains stable under all examined hydrostatic pressure conditions. 
It is worth noting that RbBaCl$_3$, RbBaBr$_3$ and RbBaI$_3$ remain thermodynamically stable up to 90 GPa, 70.4 GPa, and 49 GPa, respectively.

\begin{table}[t]
\caption{Calculated lattice constant (\AA) for RbBaX$_3$ (X = F, Cl, Br, I). The pressure is varied from 0 to 90 GPa. The lattice constants from existing literature for the RbBaF$_3$ material are also added with references. 
}
\centering\label{table:lattice_constant}
\bgroup
\def\arraystretch{1.2} 	
\begin{tabularx}{0.48\textwidth}{l @{\extracolsep{\fill}} cccc}
\hline
\hline 
\multirow{2}{*}{Materials}  & \multicolumn{4}{c}{Pressure (GPa)}\tabularnewline
\cline{2-5}
 & 0 & 30 & 60 & 90 \tabularnewline
\hline 
\multirow{2}{*}{RbBaF$_{3}$} &  4.99 & 4.47 & 4.24 & 4.09 \tabularnewline
& 4.95 \cite{RbBaF3_GW}, 5.11 \cite{RbBaF3} &  &  &  \tabularnewline
\multirow{1}{*}{RbBaCl$_{3}$} & 6.00 & 5.04 & 4.74 & 4.55 \tabularnewline
\multirow{1}{*}{RbBaBr$_{3}$} & 6.29 & 5.22 & 4.91 & 4.69  \tabularnewline
\multirow{1}{*}{RbBaI$_{3}$} & 6.72 & 5.48 & 5.14 & 4.82 \tabularnewline
\hline 
\end{tabularx}
\egroup
\end{table}

\begin{figure}[t]
\centering
\includegraphics[height=4.0in]{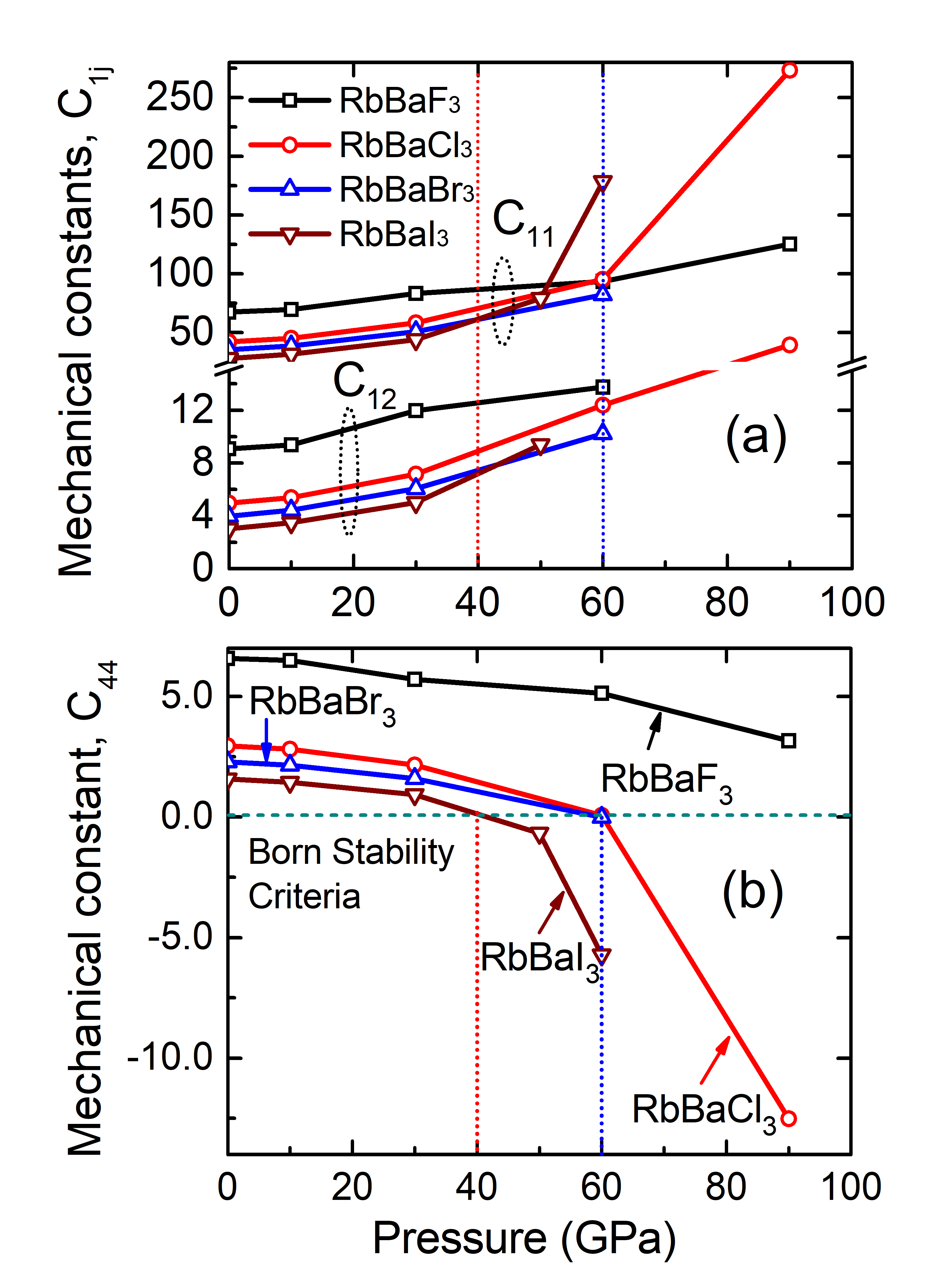}
\caption{Mechanical constants of RbBaX$_3$: (a) C$_{44}$ and (b) C$_{11}$, C$_{12}$. $C_{44}>0$ indicates stable region defined by the Born stability.}
\label{fig:RbBaX3_Cij}
\end{figure}

\begin{table*}[t]
\caption{Calculated elastic constants ($C_{11}$, $C_{12}$, $C_{44}$), Cauchy Parameter ($C_{12}-C_{44}$), bulk modulus (B), Young Modulus (Y), shear modulus (G), Poisson's Ratio ($\nu$), Pugh's ratio (B/G) for RbBaX$_3$ (X = F, Cl, Br, I). The pressure is varied up to the mechanical stability limit for each material.}
\centering\label{table:mechanical}
\bgroup
\def\arraystretch{1.0} 	
\begin{tabularx}{1.0\textwidth}{cc@{\extracolsep{\fill}} cccccccccc}
\hline 
\hline 
\multirow{2}{*} {Material} &	Pressure & $C_{11}$ & $C_{12}$ & $C_{44}$ & $C_{12}-C_{44}$ & B & Y & G & $\nu$ & B/G \tabularnewline
 &	GPa &  GPa & GPa & GPa & GPa &  GPa & GPa & GPa & & \tabularnewline
\hline 
\multirow{4}{*}{RbBaF$_{3}$} & 0 &	67.41 & 9.07 & 6.58 & 2.49 &
28.52	& 32.67	& 12.57	& 0.29	& 2.27
\tabularnewline
& 30 & 83.30 &	11.98 &	5.70	& 6.27
&	35.75 & 34.71	& 13.14 & 0.32 & 2.72
\tabularnewline
& 60 & 93.50 &	13.76	& 5.13	& 8.62 &
	40.34 & 35.76 & 13.45 & 0.33 & 2.99
\tabularnewline
& 90 &	12.54	& 19.98	& 3.16 &	16.82
 & 55.12 & 37.64 &	14.02 &	0.34 &	3.93
\tabularnewline
\hline 
\multirow{3}{*}{RbBaCl$_{3}$} & 0 & 42.03 & 4.97 & 2.95	 & 2.02 & 17.32 & 17.85 &	6.81 &	0.31 &	2.54
\tabularnewline
& 30 & 58.28 &	7.16	& 2.16 &	5.01
& 24.20 & 19.78 & 7.46	& 0.33	& 3.37
\tabularnewline
& 60 & 95.21	& 12.38	& 0.07	& 12.31
& 39.99 & 22.06 & 8.36 & 0.32 & 4.78
\tabularnewline
\hline 
\multirow{3}{*}{RbBaBr$_{3}$} & 0	& 35.29 & 	3.98 &	2.29 &	1.69 & 14.42	& 14.55 &	5.55 &	0.31 &	2.59
\tabularnewline
& 30 &	50.75 & 6.07	& 1.59	& 4.47 & 
 20.96 & 16.49 & 6.22 &	0.33	& 3.37
\tabularnewline
& 60 &	82.16	& 10.22 &	-0.02	& 10.23
 & 34.19 &	18.87	& 7.18 &	0.32 &	4.77
\tabularnewline
\hline 
\multirow{2}{*}{RbBaI$_{3}$} & 0 &	27.88 &	3.03 &	1.58 &	1.45
 & 11.31 & 10.95 &	4.17	& 0.31 &	2.71
\tabularnewline
& 30 & 43.94 &	5.01 &	0.93 &	4.08 &
	17.98 &	13.03 &	4.92 &	0.32 &	3.65
\tabularnewline
\hline 
\end{tabularx}
\egroup
\end{table*}

\subsection{Mechanical properties}
To investigate the mechanical nature of these halide perovskites,
we have calculated the elastic constants $C_{ij}$.
For crystals with cubic symmetry, the three independent elastic constants are C$_{11}$, C$_{12}$, and C$_{44}$.
C$_{11}$ is a metric quantifying the resistance against linear compressive force, whereas C$_{12}$ and C$_{44}$ are metrics indicating material resistance due to shape change.
Figure \ref{fig:RbBaX3_Cij} shows the response of the elastic constants for different external pressure conditions. 
To assess the stability of the cubic phases of RbBaX$_3$ under hydrostatic pressures, 
we applied the Born stability criterion \cite{stiffness_born,SADDIQUE2022106345}, which requires $C_{11} - C_{12} > 0, \, C_{11} + 2 C_{12} > 0, \, C_{44} > 0$.
As shown in Fig. \ref{fig:RbBaX3_Cij}(a), under all examined pressure conditions, for all structures, C$_{11} > C_{22} > 0$, indicating that all Rb-based structures meet the criteria for $C_{11} - C_{12} > 0$ and $C_{11} + 2 C_{12} > 0$ conditions. 
Therefore, C$_{44}$ is the only parameter determining the mechanical stability of these structures. 
The pressure-dependent C$_{44}$ values are presented in Fig.  \ref{fig:RbBaX3_Cij}(b). 
As seen, $C_{44}$ always remains positive for RbBaF$_3$, under all examined pressure conditions.
This is attributed to the shorter bond length of fluoride atoms in RbBaF$_3$, and is consistent with the thermodynamic stability results presented in Fig. \ref{fig:lattice_vs_pressure}(b).
However, the stability of the cubic phase with pressure diminishes for the other halides under high pressures.
The RbBaCl$_3$ and RbBaBr$_3$ are stable up to 60 GPa pressure, whereas the RbBaI$_3$ are stable only up to 40 GPa.
Note that, the mechanical constants $C_{ij}$ (j = 1, 2) exhibit a relatively monotonic change with pressure. 
C$_{11}$ is most affected by the pressure-induced changes in the lattice constant, whereas the variations in C$_{12}$ and C$_{44}$ are small and gradual. 
A relatively smaller value of C$_{44}$ is an indicator that these RbBaX$_3$ perovskites have higher shearability and low hardness, which typically results in better machinability of the materials \cite{hadi2022dft}.
Compared with the thermodynamic stability results presented in Fig. \ref{fig:lattice_vs_pressure}(b), it is evident that the mechanical stability rather determines the true upper bound of the critical pressure that these materials can withstand. 
Table \ref{table:mechanical} summarizes all the extracted mechanical constants for various hydrostatic pressure conditions.

We calculated the bulk modulus (B), Young Modulus (Y), shear modulus (G), Poisson's Ratio ($\nu$), and Pugh's ratio (B/G) from the C$_{ij}$ elastic constants
Using the Voigt-Reuss-Hill formalism \cite{hill1952elastic}. 
Table \ref{table:mechanical} presents a comprehensive list of these mechanical parameters for this class of perovskite. 
The maximum pressure considered for each structure in the table corresponds to the critical pressure at which the cubic phases remain mechanically stable. 
Note that, as the pressure increases, B, G, and Y for all structures generally rise due to the reduction in the inter-atomic bond length under pressure.
We compared our calculated results with those from an analytical model
\cite{guo2022atomistic} for the bulk modulus of cubic RbBaCl$_3$, RbBaBr$_3$ and RbBaI$_3$ at ambient pressure. 
The analytical model estimates values of 17.76, 15.21, 12.36 GPa, respectively, which agrees well with our DFT-calculated values of 17.32, 14.42 and 11.31 GPa. 
The bulk modulus for RbBaX$_3$ is relatively lower when compared to that of other classes of perovskites, such as the ABO$_3$, ABS$_3$, and ABSe$_3$ structures \cite{guo2022atomistic}, which indicates better softness, lower resistance to plastic deformation and weaker stiffness in RbBaX$_3$, respectively.
However, the bulk modulus of RbBaX$_3$ falls within the mid-range values compared to that for other ABX$_3$ members.
The RbBaF$_3$ shows the highest values of B, Y, G values among the four Rb-Ba halides, but the values decrease on average by 43\%, 54\%, 64\%, when F is substituted by Cl, Br and I, respectively.
This trend may assist in predicting the mechanical properties of unknown halide perovskites when only the mechanical properties of one halide perovskite are known.

As seen in Table \ref{table:mechanical}, the Poisson ratio remains relatively consistent, ranging from 0.29 to 0.34, across all four materials and different pressure levels.
Having a Poisson's ratio ($\nu$) higher than 0.26 is an indicator that these materials will be ductile in nature.
In contrast, Pugh’s ratio, B/G, varies from 2.27 to 2.71, depending on the size of the halide atoms. 
As pressure increases, the Pugh’s ratio rises by 0.5\% to 1.5\% per 1 GPa change, with the most significant changes observed in RbBaI$_3$. 
A Pugh’s ratio greater than the critical limit of 1.75 can be used to distinguish between the brittle or ductile nature of a material \cite{diao2022high}.
As evident from Table \ref{table:mechanical}, the Pugh’s ratio exceeds 1.75 for all the RbBaX$_3$ materials, suggesting that these materials are ductile at ambient and under high pressure conditions. 
Furthermore, a positive value of the Cauchy Parameter, $C_{12}-C_{44}$, further supports the ductility of the cubic phases of RbBaX$_3$ materials.

\begin{figure*}[t]
\centering
\includegraphics[height=4.0in]{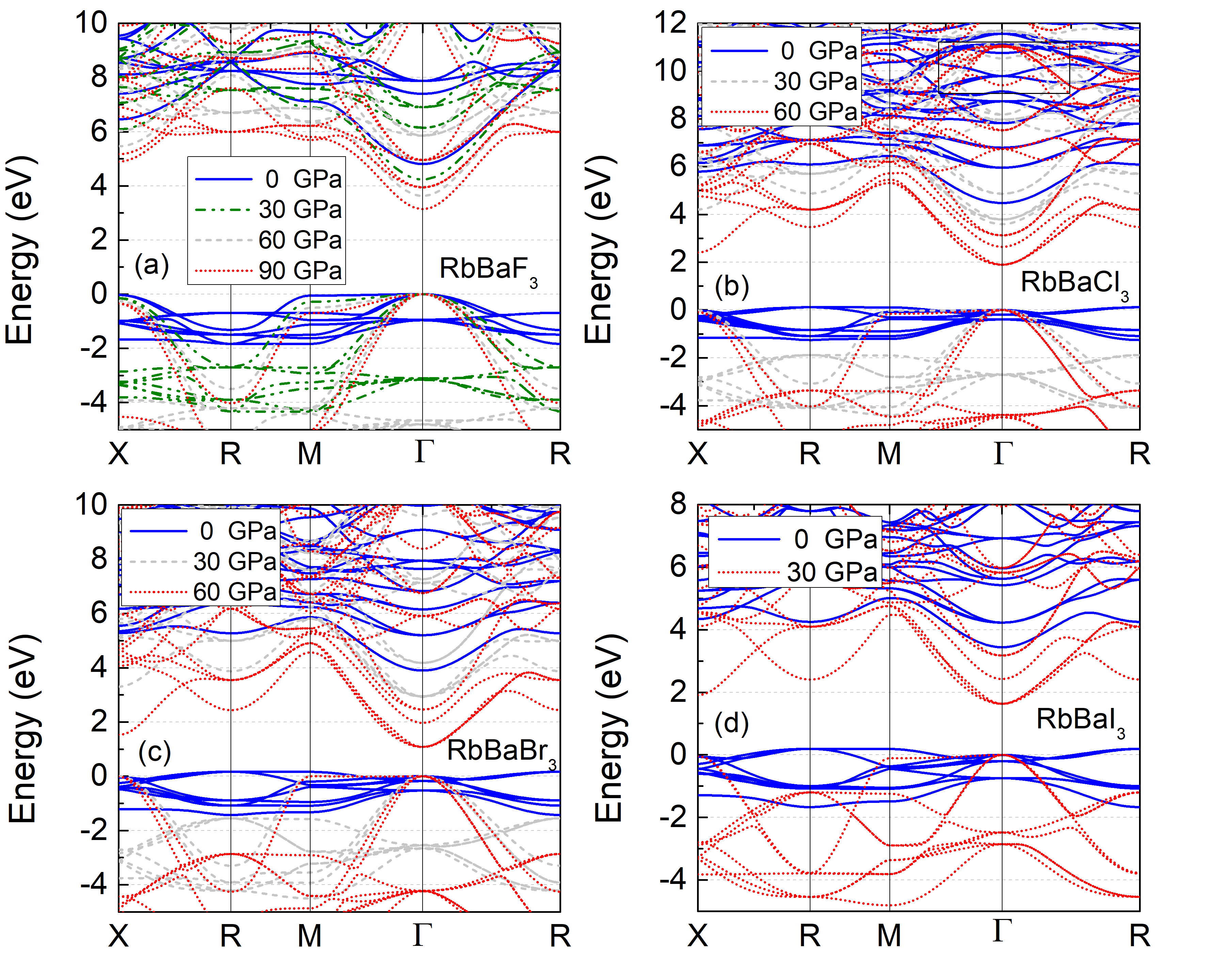}
\caption{Calculated band profile for (a) RbBaF$_3$, (b) RbBaCl$_3$, (c) RbBaBr$_3$, (d) RbBaI$_3$. The bands are calculated for different hydrostatic pressure conditions. The band structure is shifted in energy scale so that the valance band maxima at $\Gamma$ point is always at 0 eV. The high symmetry points (0,0,0), (0.5,0,0), (0.5,0.5,0), (0.5,0.5,-0.5) are denoted by $\Gamma$, X, M, R, respectfully. }
\label{fig:RbBaX3_band}
\end{figure*}

\begin{figure*}[t]
\centering
\includegraphics[height=3.2in]{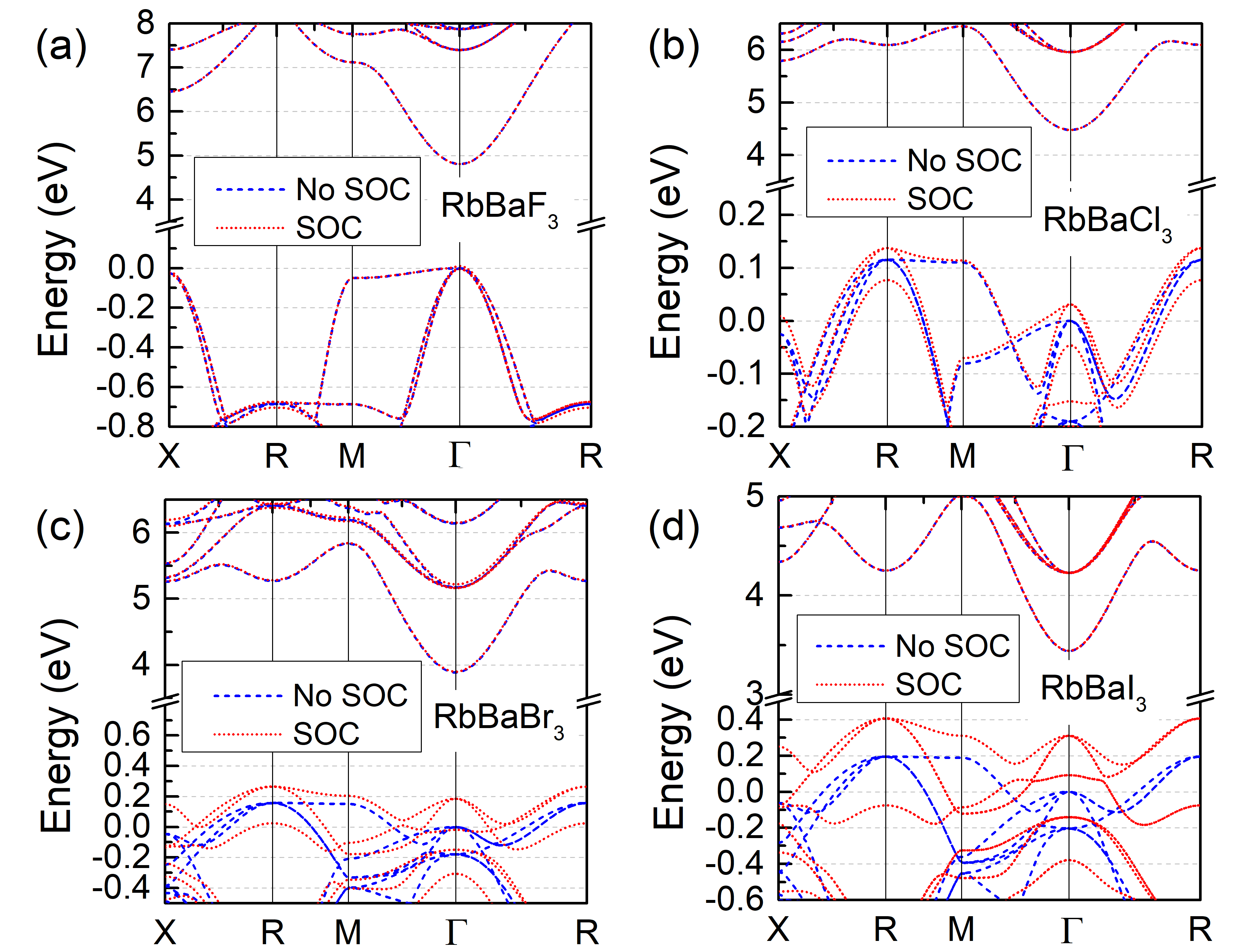}
\caption{Simulated PBE-GGA band profile in the presence of spin-orbit coupling, for (a) RbBaF$_3$, (b) RbBaCl$_3$, (c) RbBaBr$_3$, (d) RbBaI$_3$, at ambient pressure condition. For better visibility, the vertical energy scale is broken to enlarge the changes seen in the valance band regions.}
\label{fig:RbBaX3_SOC}
\end{figure*}

\begin{figure*}[t]
\centering
\includegraphics[height=3.4in]{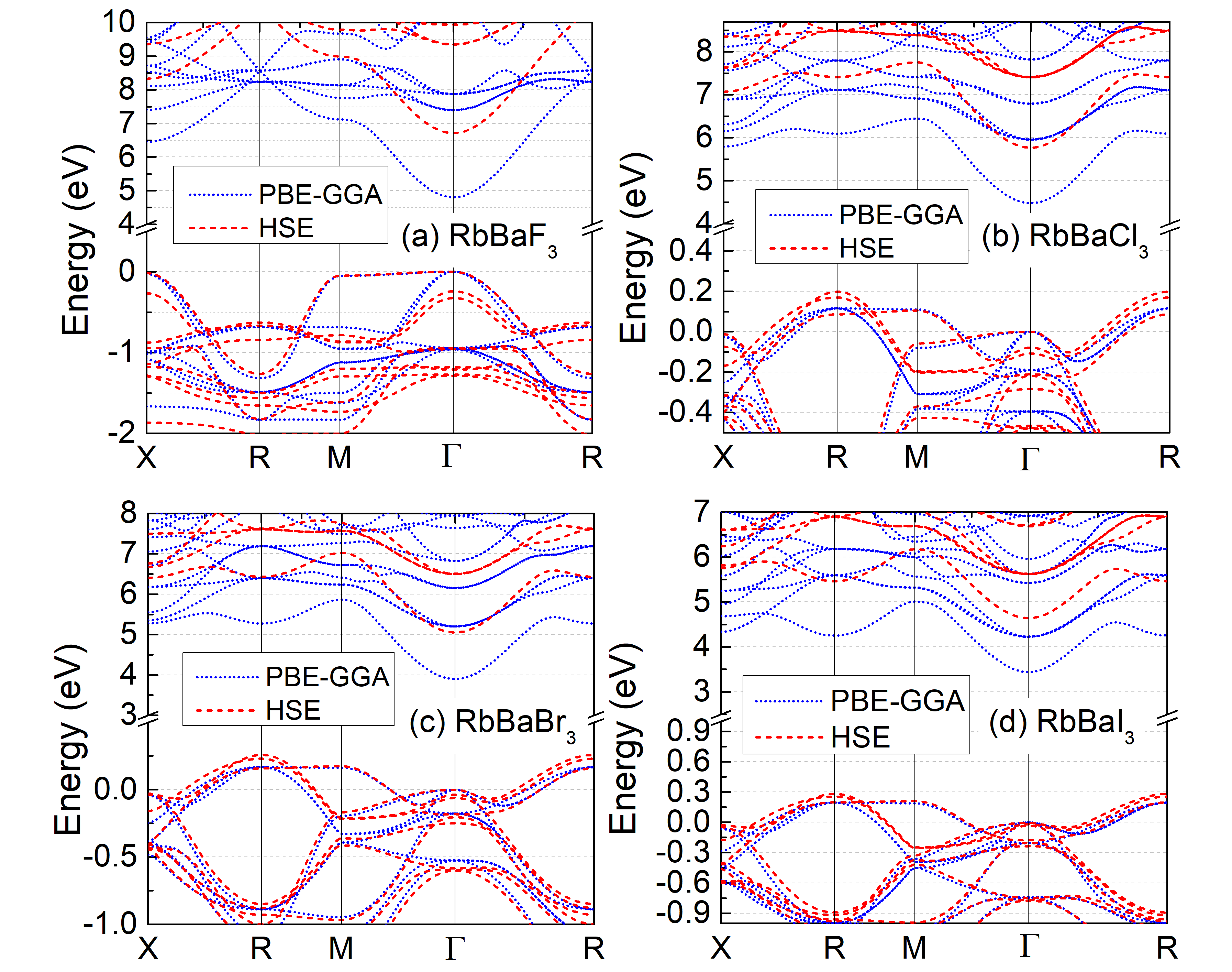}
\caption{PBE and HSE band profile for (a) RbBaF$_3$, (b) RbBaCl$_3$, (c) RbBaBr$_3$, (d) RbBaI$_3$, at ambient pressure condition. The band structure is shifted in energy scale so that the valance band maxima at $\Gamma$ point is always at 0 eV. For better visibility, the vertical energy scale is broken to enlarge the changes seen in the valance band regions.}
\label{fig:RbBaX3_hse}
\end{figure*}

\begin{figure*}[t]
\centering
\includegraphics[height=2.8in]{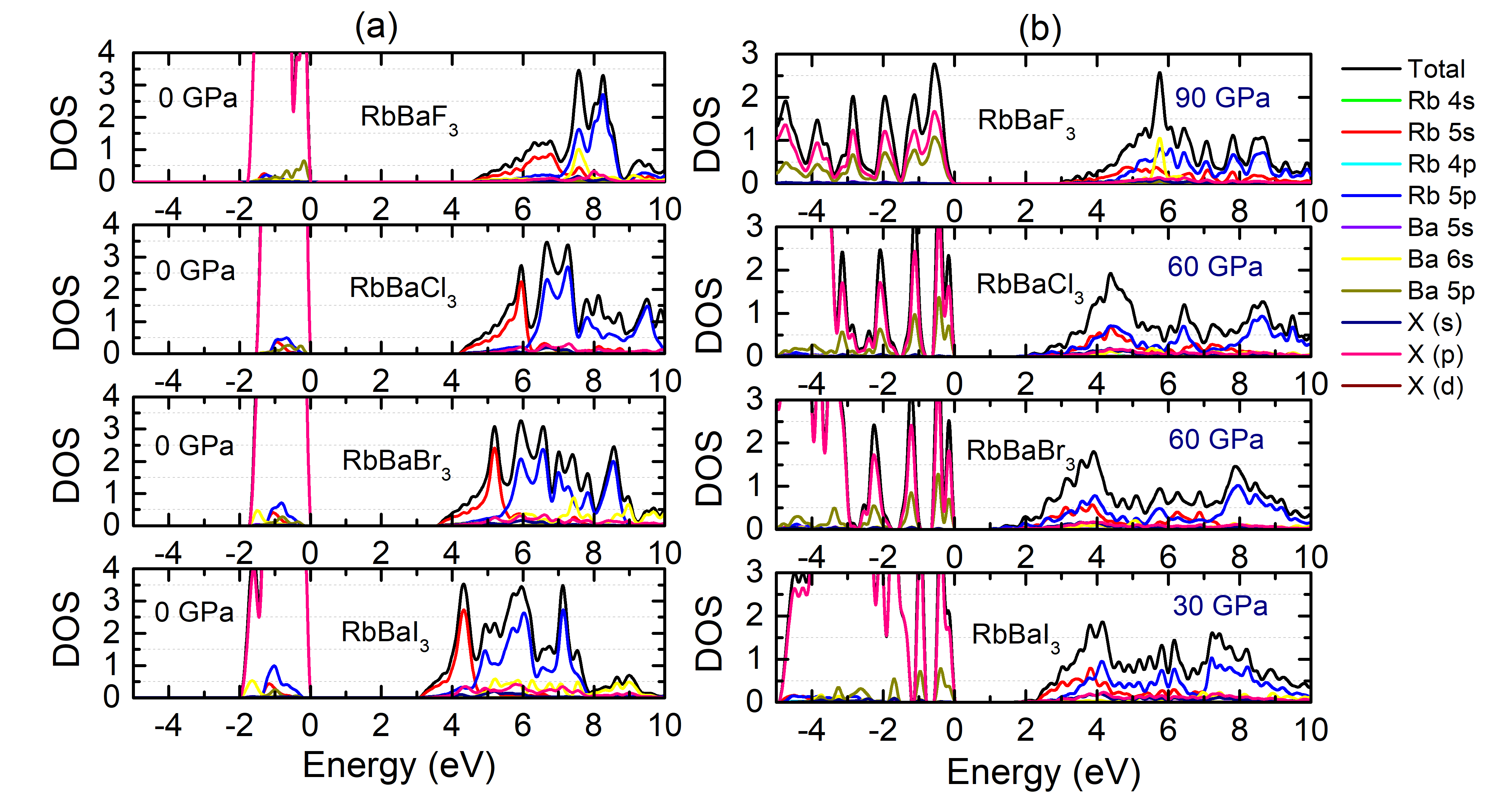}
\caption{Total density of states (DOS) and the orbital-decomposed DOS of RbBaF$_3$, RbBaCl$_3$, RbBaBr$_3$, and RbBaI$_3$: (a) for 0 GPa, (b) for the pressure close to the critical pressure condition of respective RbBaX$_3$ materials. The energy scale is shifted laterally to place the valance band maxima always at 0 eV.}
\label{fig:RbBaX3_DOS}
\end{figure*}

\subsection{Electronic properties and band structure}
Given the notable light absorption of halide perovskites and their applications in optoelectronic devices such as solar cells, it is important to understand the electronic properties of RbBaX$_3$, as the band gap is directly related to the light conversion efficiency \cite{PCE_Bandgap,review_bandgap_perovskites}.
To identify whether the RbBaX$_3$ materials exhibit metallic or insulating characteristics, we have calculated the band structure for all of the RbBaX$_{3}$ halide perovskites.
Figure \ref{fig:RbBaX3_band} (a)-(d) shows the band profiles under different pressure conditions.
The band structures are shifted along the energy scale so that the valance band maxima (VBM) at $\Gamma$ point is always set at 0 eV, making comparisons among the band dispersions of the perovskites easier. 

Figure \ref{fig:RbBaX3_band} (a) shows the electronic band structure for RbBaF$_3$. 
At no-pressure conditions, it is an insulator with 4.8 eV band gap based on PBE-GGA calculations.  
This agrees with prior theoretical estimates which  reported band gaps of 4.9 eV and 4.46 eV, using the GGA and LDA pseudopotentials, respectively \cite{RbBaF3, Nishimatsu_2002}.
The discrepancy with the LDA results is expected since LDA typically "over-binds", leading to smaller lattice constants and band gaps compared to the theoretical PBE-GGA results and the experimental values \cite{LDA_overbinds}. 
One should note that the valance band edge at the $X$ symmetry point of the Brillouin zone (BZ) is lower than the valance band edge at the $\Gamma$ point. 
This is more evident in Fig. \ref{fig:RbBaX3_SOC}(a), where a zoomed view of the valance bands are presented.
As seen, the VBM at X and M are 25.7 meV and 49.4 meV below the VBM at $\Gamma$, respectively.
Since the $\Gamma$-$\Gamma$ shows the lowest transition energy, we conclude that RbBaF$_3$ is a direct band gap insulator. 
Our conclusion contradicts with earlier LDA \cite{Nishimatsu_2002} and GGA \cite{RbBaF3} reports, but agrees with more accurate DFT simulation based on LDA+GW formalism \cite{RbBaF3_GW}.
Note that, the valance-band energy differences of $\Gamma$-X and $\Gamma$-M are just 1$\sim$2 $K_{B}T$ ($K_{B}$ = Boltzmann constant, $T$ = ambient temperature) and thus RbBaF$_3$ may transition to an indirect band gap insulator due to minor external stimuli, e.g., spin interaction, pressure, temperature etc.
Since relaxed cubic RbBaF$_3$ is a non-magnetic material, there is no impact from imposing collinear spin texture into the calculation.
However, the spin-orbit coupling (SOC) is one important factor that operates on the meV energy scale.
For instance, SOC has been demonstrated to cause a band splitting of 30 meV in fluorinated graphene when compared to the hydrogenated graphene \cite{irmer2015spin}.
Despite this, earlier \textit{ab initio} simulations of RbBaF$_3$ have so far ignored the SOC effect. 
In Fig. \ref{fig:RbBaX3_SOC}(a), 
we quantify the effect of SOC on the band structure of RbBaF$_3$. 
Our finding reveals that while the SOC does not alter the  conduction bands, it introduces a split in the doubly-degenerate valance band maxima at X, causing
the VBM(X) to shift upwards by 7 meV.
The energy shift still maintains $\Gamma$-$\Gamma$ the prominent transition path.
Hence, we conclude that the SOC effect is relatively small for RbBaF$_3$ and RbBaF$_3$ remains a stable direct band gap insulator.

Fig. \ref{fig:RbBaX3_band}(a) also shows the impact of hydrostatic pressure on the band profile and band transition. 
As seen, even under pressure up to 90 GPa, RbBaF$_3$ still exhibits direct transition with a band gap of 3.2 eV.
Unlike the SOC effect, the VBM(X) point decrease further relative to VBM($\Gamma$) edge by 140, 250, and 343 meV for pressures of 30, 60, and 90 GPa, respectively. 
The trend indicates that RbBaF$_3$ also
remains insusceptible to any pressure-induced direct-to-indirect band transitions.

RbBaCl$_3$ exhibits a relatively different behavior
as shown in Fig. \ref{fig:RbBaX3_band}(b) and \ref{fig:RbBaX3_SOC}(b).
At ambient pressure, RbBaCl$_3$ is an indirect band-gap material.
The lowest energy transition occurs between the conduction band minima at $\Gamma$ point and the valance band maxima at R point, with a band gap of 4.37 eV.
As pressure increases from 0 GPa to 30 GPa and to 60 GPa, both VBM(X) and VBM(M) moves lower than VBM($\Gamma$) point. 
This shift results in a pressure-induced transition from an indirect to a direct band gap in the chloride perovskite. 
Since an indirect band gap means more phonon channels generating heat \cite{islam2020semiconducting}, such indirect-to-direct band transition makes strained RbBaCl$_3$ more promising for efficient optoelectronic devices than its unstrained counterpart.  
For the hydrostatic pressure of 30 GPa and 60 GPa, the band gaps in RbBaCl$_3$ are 3.6 eV (direct) and 1.9 eV (direct), respectively.
It is worth mentioning that as the pressure approaches 90 GPa, the RbBaCl$_3$ also shows a semiconductor-to-metal transition.
However, as discussed previously, this metallic transition is not observed in a mechanically stable phase and thus we did not present the band structure under 90 GPa pressure. 
We infer that RbBaCl$_3$ is essentially an indirect insulator with the potential for pressure-induced direct transition.
Figure \ref{fig:RbBaX3_SOC} (b) presents the band structure at ambient pressure considering the SOC effect.
The valance band edge at $\Gamma$ is more significantly affected by SOC (31 meV shift) than the VBM at R point (22.2 meV shift). 
This suggests that, unlike RbBaF$_3$, SOC effect in RbBaCl$_3$ actively reduces the energy differences between direct and indirect transitions.
Consequently, RbBaCl$_3$ can more readily undergo a transition to direct band gap under external pressure.

The other two materials, RbBaBr$_3$ and RbBaI$_3$, show similar band evolution in response to external pressure. 
RbBaBr$_3$ shows an indirect $\Gamma$-R band gap of 3.7 eV at no-pressure condition.
As the pressure increases from 0 GPa to 30 GPa and to 60 GPa, it remains an indirect material, but the indirect band transition shifts from $\Gamma$-R to $\Gamma$-M.
Similarly, RbBaI$_3$ starts with an indirect $\Gamma$-R band gap of 3.24 eV.
Like RbBaCl$_3$, RbBaI$_3$ becomes a direct band gap material at 30 GPa pressure.
For both Br and I, the effect of SOC is most significant. 
As shown in Fig. \ref{fig:RbBaX3_SOC}(c)-(d), SOC increases the VBM($\Gamma$) energy of RbBaBr$_3$ and  RbBaI$_3$ by 185 meV and 311 meV, respectively.
Similar to RbBaCl$_3$, the SOC effect reduces the energy differences between the VBM at $\Gamma$ and R, thereby decreasing the direct $\Gamma$-$\Gamma$ band gap more significantly than the indirect $\Gamma$-R band gap.

It is well-established that PBE-GGA theory often underestimates the true band gap of a material \cite{sun2016accurate}, although the qualitative conclusions derived from PBE-GGA generally remain valid.
To obtain a more realistic band gap for RbBaX$_3$,
we have simulated the band structure with the hybrid functional correction.
As shown in Fig. \ref{fig:RbBaX3_hse}, the HSE simulation introduces only minor adjustments on the valance band profiles, with the VBM at R shifting upward by 50 to 80 meV.
However, the most drastic change occurs at the conduction bands edges.
The application of HSE results in
an increased band gap from the calculated PBE-GGA values to 6.71 eV (direct) for RbBaF$_3$, 5.57 eV (indirect) for RbBaCl$_3$, 4.79 eV (indirect) for RbBaBr$_3$, and 4.36 eV (indirect) for RbBaI$_3$.
Table \ref{table:bandgap} summarizes the band transition energies for all four materials, with PBE, PBE+SOC and HSE theory.
It also highlights the band transitions from indirect to direct for RbBaCl$_3$ and RbBaI$_3$ under different hydrostatic pressure conditions. 
Importantly, the nature of the band transitions (direct vs. indirect) with PBE+SOC and HSE theory remain unchanged as in PBE-GGA theory.
The primary difference is only the magnitude of the band gap provided by the HSE.

\begin{table*}[t]
\caption{Band-to-band transition energy (in eV), from CBM (at $\Gamma$) to different high symmetry locations of the VBM ($\Gamma$, X, M, R), for PBE, PBE+SOC and HSE level of theory. The direct band gaps are indicated by $*$, and indirect band gaps are denoted by \textdagger.}
\centering\label{table:bandgap}
\bgroup
\def\arraystretch{1.0} 	
\resizebox{\textwidth}{!}{\begin{tabularx}{0.98\textwidth}{c|c|cccc|cccc|cccc}
\hline 
\hline 
\multirow{2}{*} {Material} & Pressure & \multicolumn{4}{c|}{PBE} & \multicolumn{4}{c|}{PBE + SOC} & \multicolumn{4}{c}{HSE} \tabularnewline
\cline{3-14}
& GPa & $\Gamma$-$\Gamma$ & $\Gamma$-X &  $\Gamma$-M & $\Gamma$-R & $\Gamma$-$\Gamma$ & $\Gamma$-X & $\Gamma$-M & $\Gamma$-R & $\Gamma$-$\Gamma$ & $\Gamma$-X & $\Gamma$-M & $\Gamma$-R \tabularnewline
\hline 
\multirow{4}{*}{RbBaF$_{3}$} 
& 0 & 4.80{}\textsuperscript* & 4.83 & 4.85 & 5.49 & 4.79{}\textsuperscript* & 4.83 & 4.85 & 5.48 & 6.71{}\textsuperscript* & 6.73 & 6.76 & 7.34
\tabularnewline
& 30 & 4.24{}\textsuperscript* & 4.37 & 4.52 & 6.92  & & & & & & & &
\tabularnewline
& 60 & 3.63{}\textsuperscript* & 3.88 & 4.13 & 7.12  & & & & & & & & 
\tabularnewline
& 90 & 3.14{}\textsuperscript* & 3.48 & 3.83 & 7.16   & & & & & & & &
\tabularnewline
\hline 
\multirow{3}{*}{RbBaCl$_{3}$} 
& 0 & 4.48 & 4.50 & 4.56 & 4.37{}\textsuperscript\textdagger & 4.37 & 4.39 & 4.54 & 4.25{}\textsuperscript\textdagger & 5.77 & 5.78 & 5.82 & 5.57{}\textsuperscript\textdagger
\tabularnewline
& 30 & 3.59{}\textsuperscript* & 4.42 & 3.62 & 5.47  & & & & & & & &
\tabularnewline
& 60 & 1.89{}\textsuperscript* & 1.98 & 2.05 & 5.25   & & & & & & & &
\tabularnewline
\hline 
\multirow{3}{*}{RbBaBr$_{3}$} 
& 0 & 3.9 & 3.94 & 3.74 & 3.73{}\textsuperscript\textdagger & 3.71 &	3.74 &	3.68 &	3.62{}\textsuperscript\textdagger &	5.05 &	5.08 &	4.88 &	4.79{}\textsuperscript\textdagger
\tabularnewline
& 30 & 2.94 &	2.94 &	2.78{}\textsuperscript\textdagger &	4.49 & & & & & & & &
\tabularnewline
& 60 & 1.08 &	1.09 &	1.06{}\textsuperscript\textdagger &	3.95 & & & & & & & &
\tabularnewline
\hline 
\multirow{2}{*}{RbBaI$_{3}$} 
& 0 & 3.44 &	3.50 &	3.25 &	3.24{}\textsuperscript\textdagger &	3.09 &	3.69 &	3.12 &	3.03{}\textsuperscript\textdagger &	4.64 &	4.67 &	4.43 &	4.36{}\textsuperscript\textdagger
\tabularnewline
& 30 & 1.63{}\textsuperscript* &	1.69 &	1.74 &	2.83 & & & & & & & &
\tabularnewline
\hline 
\end{tabularx}}
\egroup
\end{table*}

\subsection{Density of states}
To elucidate the contributions of the atomic orbitals on the band edges, we calculate the density of states (DOS) for all the RbBaX$_3$ perovskites, as shown in Fig. \ref{fig:RbBaX3_DOS}.
It is quite apparent that the valance band (E<0) is controlled by the outermost \textit{p}-orbital of the respective halogen atoms. 
On the other hand, the conduction band is dominated mostly by the Rb 5\textit{s}-orbital.
For the energy level much higher than the conduction band minima, the contribution switches from Rb 5\textit{s} to 5\textit{p}.
The contributions of the other orbitals on the band edges are quite negligible at no-pressure condition.
The DOS also explains why the spin-orbit coupling causes splits in the valance bands, but does not impact the conduction bands much. 
Since valance band is controlled by halogen \textit{p}-orbital, the larger the halogen atom is, the greater is the spin-induced band movement.
On the other hand, since the SOC should not introduce any band splitting for the \textit{s}-orbital,   
the conduction bands near the CBM remain unchanged due to SOC.
Only when the Rb-5\textit{p} orbital contribution will start to dominate deep in the conduction bands, then the SOC will start showing the spin-induced band non-degeneracy on those regions of the conduction band.

Figure \ref{fig:RbBaX3_DOS} also shows the density of states at higher pressure conditions, specifically when the materials are approaching their critical stability limit. 
The density of states from the valance bands starts to spread out more with higher pressure, resulting in a reduction of the strength of the DOS, while exhibiting more peaks from the spread-out phenomena.
This agrees with the pressure-induced valance-band movement behavior seen in Fig. \ref{fig:RbBaX3_band}.
The percentage contribution of Ba 5\textit{p}-orbital compared to halogen-\textit{p} orbital also increases with high pressure, indicating that SOC effect from Ba, albeit small, will add to the existing SOC effect from halogen.
For the conduction bands, the pressure introduces more hybridization of Rb-5\textit{s} and Rb-5\textit{p} orbitals, as evident from the DOS levels above CBM seen in Fig. \ref{fig:RbBaX3_DOS}(b).
This means that, unlike the 0-GPa case, the effect of SOC will start to dominate the lower-level conduction bands much early, when strain is present in the system.

\begin{figure*}[t]
\centering
\includegraphics[height=3.5in]{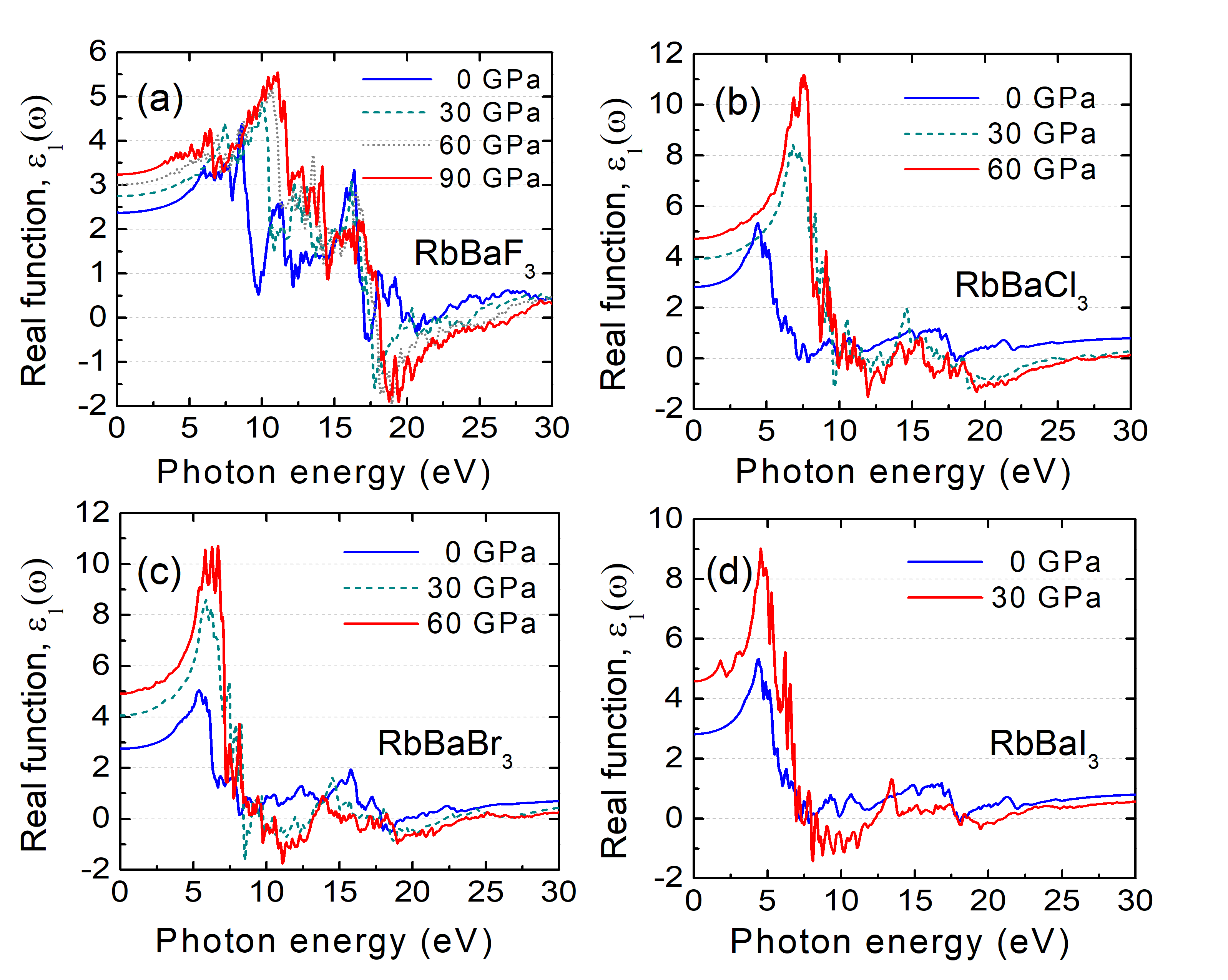}
\caption{Real part of the dielectric function, $\epsilon_1(\omega)$ for (a) RbBaF$_3$, (b) RbBaCl$_3$, (c) RbBaBr$_3$, (d) RbBaI$_3$. Both ambient pressure and pressure close to the critical conditions, at a step of 30 GPA, have been considered.}
\label{fig:RbBaX3_epsilon_real}
\end{figure*}

\begin{figure*}[t]
\centering
\includegraphics[height=3.5in]{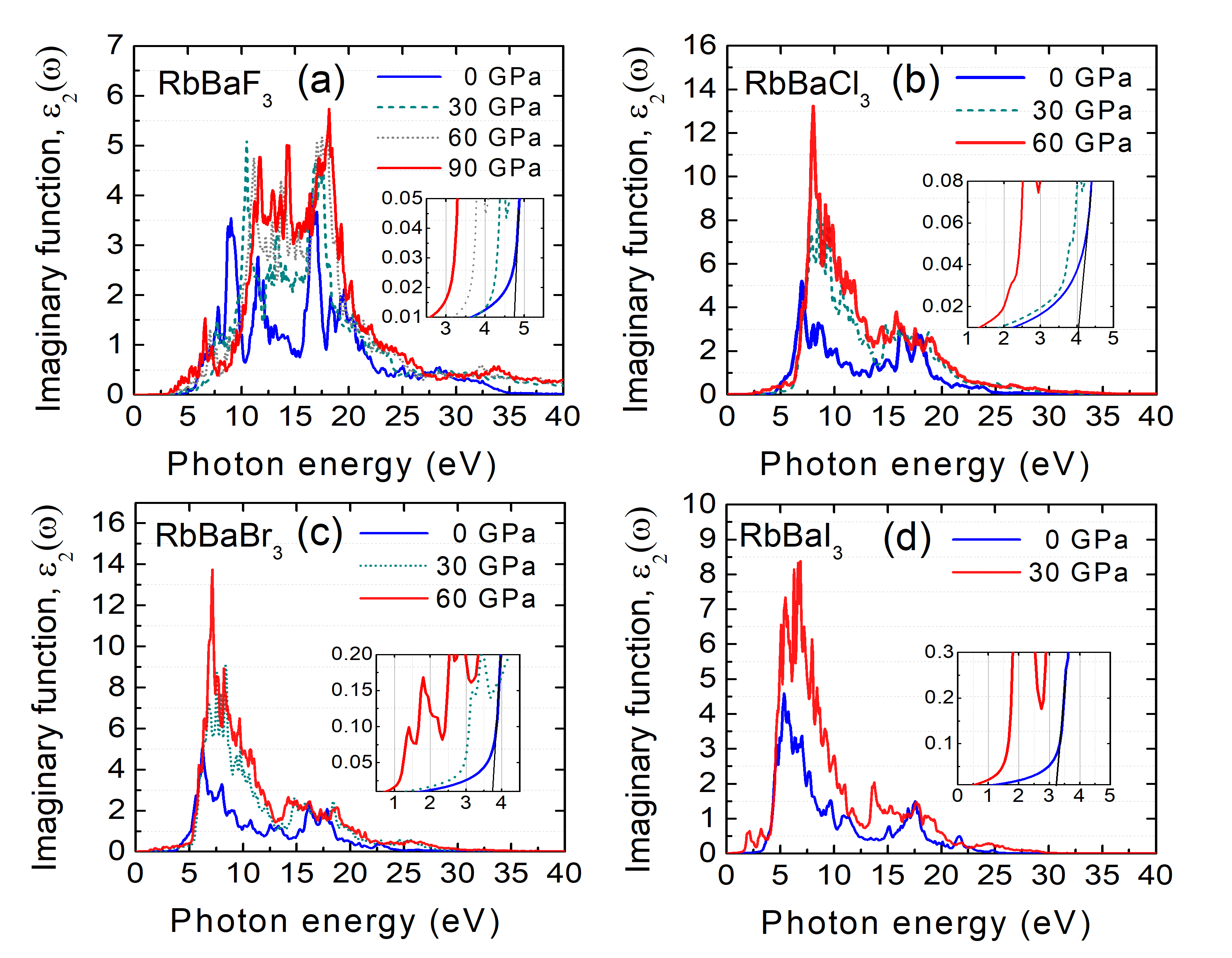}
\caption{Imaginary part of the dielectric function, $\epsilon_2(\omega)$ for RbBaX$_3$. Both ambient pressure and pressure close to the critical conditions, at a step of 30 GPA, have been considered.}
\label{fig:RbBaX3_epsilon_imag}
\end{figure*}

\begin{figure}[t]
\centering
\includegraphics[height=2.5in]{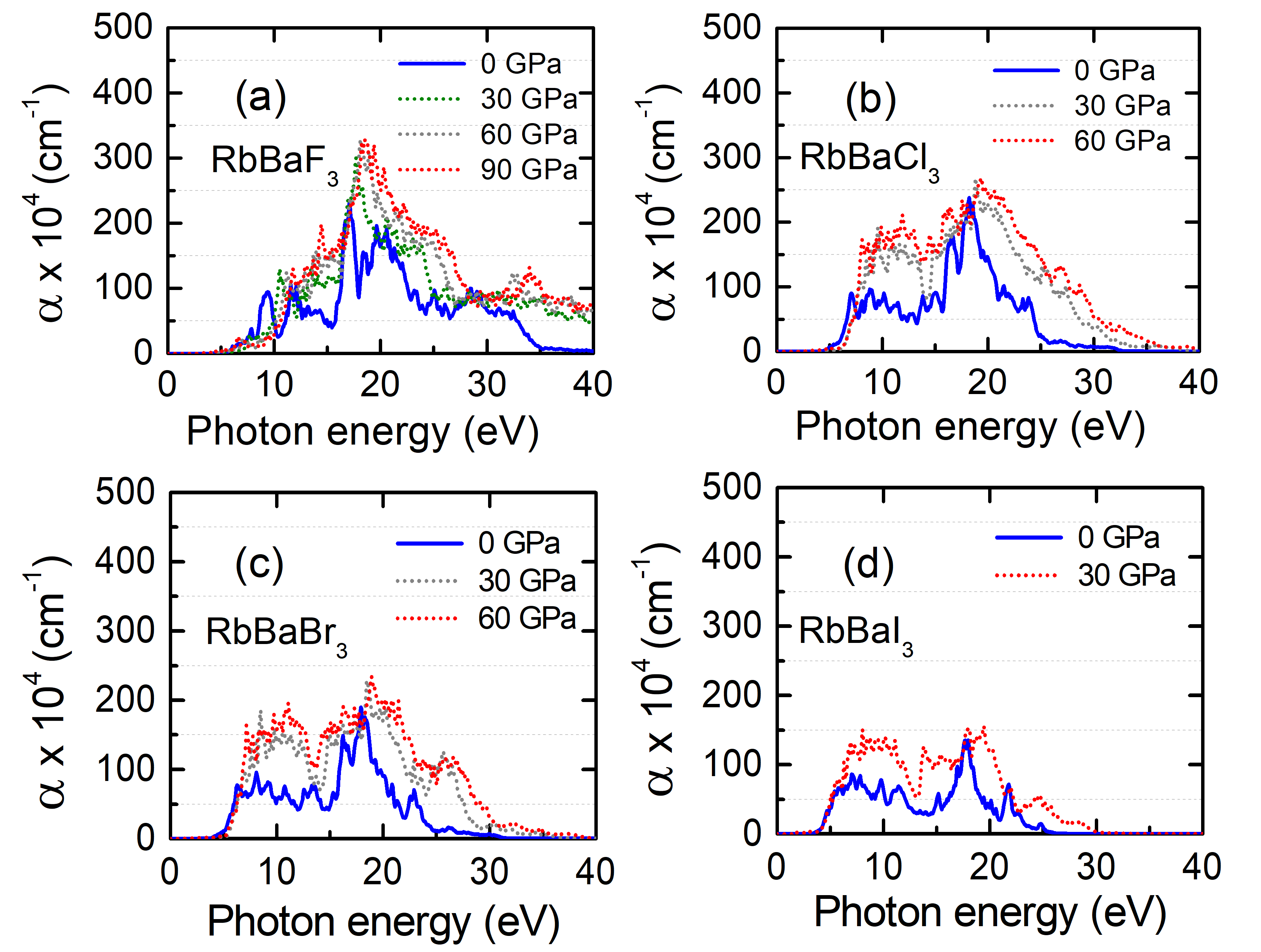}
\caption{Frequency-dependent absorption coefficient $\alpha$ (in cm$^{-1}$) for (a) RbBaF$_3$, (b) RbBaCl$_3$, (c) RbBaBr$_3$, (d) RbBaI$_3$. Both ambient pressure and pressure close to the critical conditions, at a step of 30 GPA, have been considered.}
\label{fig:RbBaX3_alpha}
\end{figure}

\begin{figure}[t]
\includegraphics[height=2.5in]{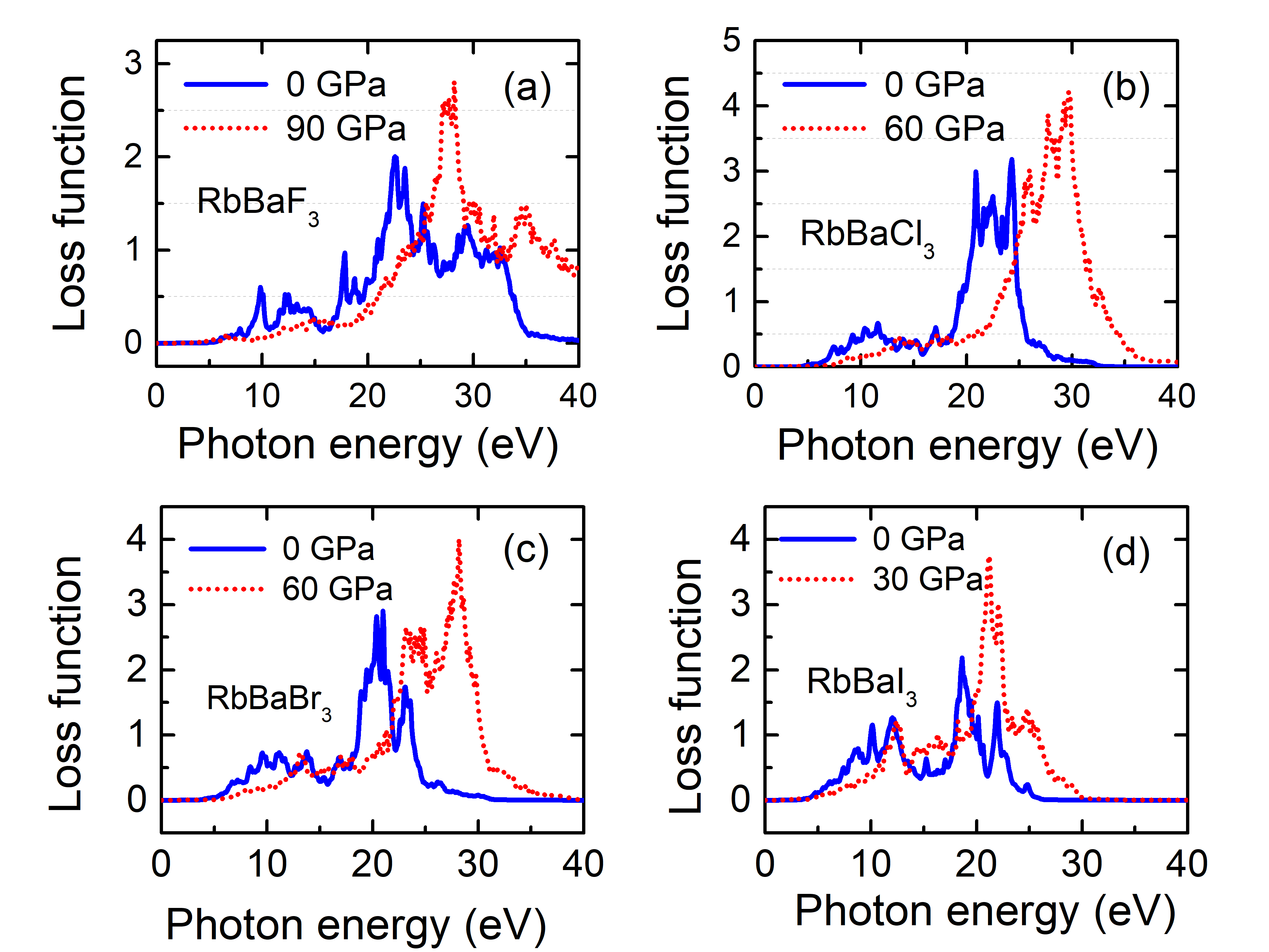}
\centering
\caption{Frequency-dependent loss function for (a) RbBaF$_3$, (b) RbBaCl$_3$, (c) RbBaBr$_3$, (d) RbBaI$_3$. Both ambient pressure and pressure close to the critical conditions have been considered.}
\label{fig:loss}
\end{figure}

\begin{figure}[t]
\includegraphics[height=2.5in]{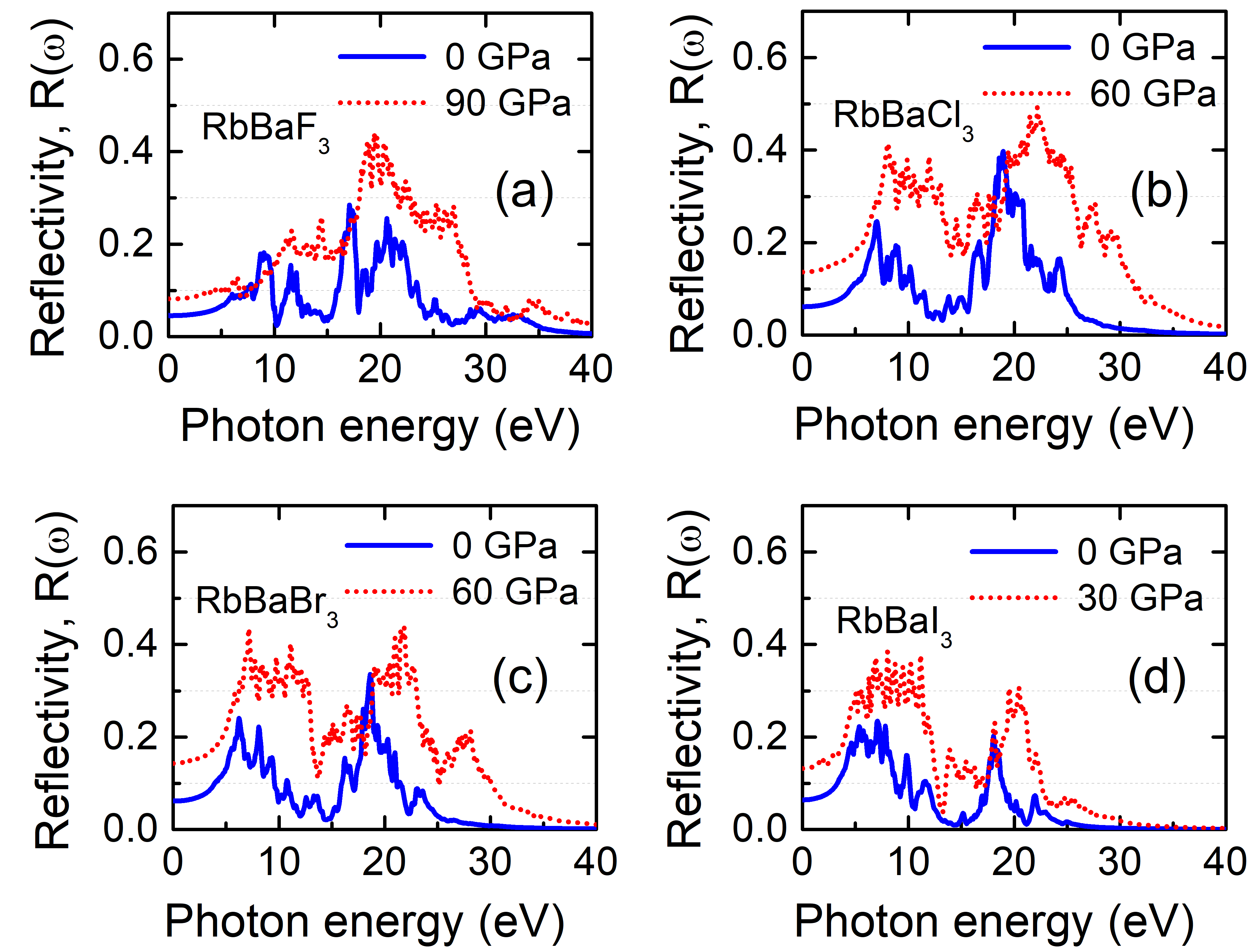}
\centering
\caption{Frequency-dependent reflectivity R($\omega$) of cubic phases of RbBaX$_3$: (a) RbBaF$_3$, (b) RbBaCl$_3$, (c) RbBaBr$_3$, (d) RbBaI$_3$. Both ambient pressure and pressure close to the critical conditions have been considered.}
\label{fig:reflectivity}
\end{figure}

\subsection{Optical Properties}
The optical properties across various photon energies offer insights into the electron-photon interaction in RbBaX$_3$ perovskites, which is closely linked to the band structure discussed previously.
The complex dielectric function $\epsilon(\omega)$, which describes the linear photovoltaic response to electromagnetic radiation, is a key indicator of material's spectral characteristics \cite{dielectric_band, optical_band, CsPbI3}. 
Figures \ref{fig:RbBaX3_epsilon_real}-\ref{fig:RbBaX3_epsilon_imag} show the effect of hydrostatic pressure on the real and imaginary components of the dielectric function $\epsilon(\omega)$, derived from the PBE-GGA calculations. 
For the four halides, the analyses consider only the pressures up to the level where each material remains mechanically stable.
The real and imaginary components of the dielectric function characterize the effect of polarization, the inter-band transition and the absorption profile of the material.

%
%
As shown in Fig. \ref{fig:RbBaX3_epsilon_real}, the static dielectric constant, $\epsilon_1(0)$, are 2.4, 2.8, 2.75, 2.82 for RbBaF$_3$, RbBaCl$_3$, RbBaBr$_3$ and RbBaI$_3$, respectively.
These values are relatively invariant to the halogen atoms, with the exception of the fluoride case, which exhibits the lowest value of $\epsilon_1(0)$.  
However, under external pressure, the sensitivities of $\epsilon_1(0)$ varies significantly from fluoride to iodide.
RbBaF$_3$ experiences a 35\% increase in $\epsilon_1(0)$ for 90 GPa, while  RbBaI$_3$ sees nearly twice that increase (62\%) but at three times lower external pressure of 30 GPa.
This suggests that larger halide ions are more sensitive to external pressure when it comes to the static dielectric constant.
The behavior can be explained by the inverse relationship of $\epsilon_1(0)$ with the band gap energy $E_g$, as expressed by the Penn Model \cite{penn1962wave}, $\epsilon_1(0) \sim 1 + (\hbar \omega_p / E_g)^2$.
Since $E_g$ variation with pressure is less steep in fluoride than in iodide (Table \ref{table:bandgap}), the change in static dielectric constant is also less sensitive to pressure in RbBaF$_3$.  
Note that, at zero pressure, for all the examined perovskites, the real dielectric function $\epsilon_1(\omega)$ stays mostly positive ($\epsilon_1> 0$) over the photon energy ($\omega$) range.
However, as pressure increases, spectral regions with negative $\epsilon_1$ emerge, 
indicating optical regions with poor transmission, poor reflectivity and optical losses \cite{wei2022defect,zahedi2015electronic}.
For RbBaF$_3$, the negative zone starts in the deep UV region of 18 eV, whereas for the other halides, it shifts towards the lower end of the spectrum of 7 $\sim$ 10 eV.

The imaginary component of the dielectric function, $\epsilon_2(\omega)$, provides information about the light absorption and energy conversion efficiency \cite{doi:10.1021/jp410579k, hybrid_energy_conversion}.
Figure \ref{fig:RbBaX3_epsilon_imag} displays the imaginary part of the dielectric function of the RbBaX$_3$ perovskites, for both normal and pressurized conditions.
The $\epsilon_2(\omega)$ function exhibits a threshold photon energy, beyond which it starts to show significant non-zero values.
We estimate the threshold energy by taking a slope at the rising sections of $\epsilon_2$.
Insets in Fig. \ref{fig:RbBaX3_epsilon_imag} presents a closer view of the regions near such turn-on behavior and the slope line for the zero-pressure case.
For RbBaF$_3$, RbBaCl$_3$, RbBaBr$_3$, and RbBaI$_3$, the estimated threshold photon energies are 4.76 eV, 4.0 eV, 3.74 eV and 3.3 eV, respectively. 
These values directly correspond to the lowest band-to-band transition seen either at the $\Gamma-\Gamma$ or $\Gamma$ to other zone edge points in these materials.
%
%
Additionally, several peaks appear in the imaginary dielectric response. 
For example, the RbBaF$_3$ shows peaks at 6.0 eV, 7.8 eV, 9.0 eV, 11.5 eV and 16.0 eV. 
These peaks are associated with transitions between the deep states of the valance bands and conduction bands, with some peaks correlated with X-X (6.5 eV), M-M (7.1 eV), R-R (8.7 eV) transitions, as shown in the band diagram in Fig. \ref{fig:RbBaX3_band}(a).

We calculated the absorption coefficient up to 40 eV for all four perovskites. 
The results show that these materials primarily absorb in the UV range of 3 $\sim$ 40 eV,
indicating potential applications as UV absorbers, stabilizers and quenchers \cite{UV_application_perovskites}. 
As seen in Fig. \ref{fig:RbBaX3_alpha}, the absorption energy shows the highest peak of 233$\times$10$^4$ cm$^{-1}$ and 238$\times$10$^4$ cm$^{-1}$ for the RbBaF$_3$ and RbBaCl$_3$ perovskites, but it gradually reduces for the heavier halogen cases.
In all cases, the absorption starts around the onset of the band gap energy of respective perovskites.
RbBaF$_3$ shows a wide absorption spectrum in the deep UV regions with the peak absorption occurring between 15 and 25 eV range.
%
As the size of the halogen atom increases from F to I, the spread of the absorption spectrum becomes narrower along the energy scale.
However, similar distinct features with multiple peaks remain visible for all compounds.
Note that, the absorption spectrum red-shifts towards the visible light region from RbBaF$_3$ to RbBaI$_3$, due to the band gap reduction.

As the pressure increases, both the strength and the range of absorption peaks expand, with the onset of absorption shifting to lower energy ranges of 1.5 to 3 eV (visible light range).
However, the strength of absorption near the visible light is comparatively small.
The observed shift in the absorption spectra with pressure aligns with the energy band variations under pressure discussed before.
As evident from Fig. \ref{fig:RbBaX3_alpha}, under compressive pressure, all the materials demonstrate elevated UV absorption, which is 2 to 4 times higher in certain optical frequency regions, when compared with the $\alpha$ of no-pressure condition.
Such pressure-induced optical response makes the RbBaX$_3$ perovskites a potential candidate for designing devices with tunable absorption spectra, where the strain can effectively tune both the range and magnitude of UV absorption. 
Additionally, the compressed RbBaX$_3$ structure absorbs more light near the visible light spectrum than the RbBaX$_3$ at ambient pressure, making the strained RbBaX$_3$ more appealing for visible-light optoelectronic applications.

\begin{figure}[t]
\includegraphics[height=2.5in]{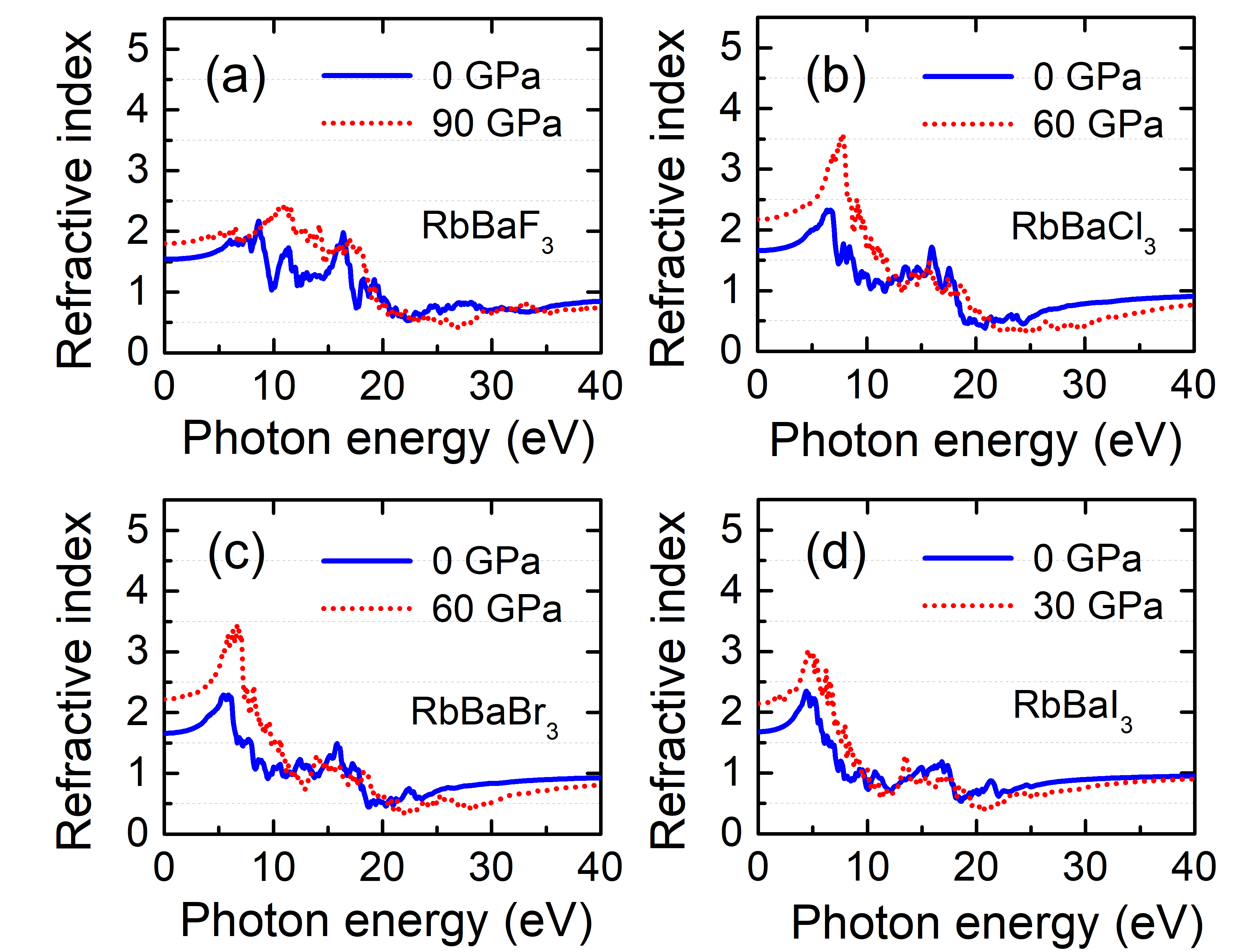}
\centering
\caption{Frequency-dependent refractive index ($n$) of cubic phases of RbBaX$_3$: (a) RbBaF$_3$, (b) RbBaCl$_3$, (c) RbBaBr$_3$, (d) RbBaI$_3$. Both ambient pressure and pressure close to the critical conditions have been considered.}
\label{fig:refractivity}
\end{figure}

We have also calculated the energy loss function for these halide perovskites.
The energy loss function, $L(\omega) = -j/\epsilon(\omega)$, quantifies the energy lost by electrons as they are traversing through the material.
The peaks observed in  $L(\omega)$, as shown in Fig. \ref{fig:loss}, indicate significant energy loss  
when the photon energy exceeds the band gap of respective perovskite.
Since there are no peaks under the 5.0 eV, one can infer that RbBaX$_3$ materials are 
optically lossless below that range.
Peak energy loss happens around 22.5 eV, 20.9 eV, 20.36 eV and 18.6 eV for the four materials, respectively. 
As shown in Figs. \ref{fig:loss} (a) - (d), with higher compressive pressure, all four materials show discernible blue shift in the optical loss towards even higher photon energies.

Figures \ref{fig:reflectivity}-\ref{fig:refractivity} present the frequency-dependent reflectivity spectrum $R (\omega)$ and refractive index $n(\omega)$ for all four RbBaX$_3$ structures. 
In the visible light spectrum (1.5-3 eV), the reflectivity R($\omega$) for all the RbBaX$_3$ perovskites is found to be around 6\%$\sim$10\%. 
Beyond the deep UV regions ($E_{ph}$ > 4.5 eV), the reflectivity starts to increase significantly from the static limit.  
Furthermore, with high pressure, the reflectivity increases 1.5 to 2 times in certain regions. 
The reflectivity response due to external pressure is higher for the perovskites with larger halogen sizes. 
However, the maximum reflectivity does not exceed 50\% for this class of materials, even with critical pressure.
The reflectivity spectrum exhibits a small red-shift of 2-3 eV from RbBaF$_3$ to RbBaI$_3$, whereas the external pressure introduces a blue shift of 2-3 eV, suggesting a material-dependent and pressure-dependent $R(\omega)$ tunability.
In case of the refractive nature of these perovskites, the static refractive indices for RbBaF$_3$, RbBaCl$_3$, RbBaBr$_3$, RbBaI$_3$ are 1.54, 1.66, 1.67, 1.68, respectively.
Under pressure, RbBaF$_3$ sees a smaller increase in the static refractive indices ($\sim$16\%), whereas the other halides see a much larger change ($\sim$30\%) in $n(\omega)$.
The refractive index metric undergoes a significant reduction for the higher-energy photons in the deep UV regions, for both the ambient condition and under external hydrostatic pressure.

\section{Conclusion}

The structural, mechanical, electronic and optical properties of RbBaX$_3$ perovskites have been investigated using \textit{ab initio} density functional theory.  
The negative formation energy indicates that these materials are thermodynamically stable.  
The calculated elastic constants and the Born stability criteria suggests that RbBaX$_3$ perovskites are mechanically stable and remain so up to certain critical pressures.
Additionally, the calculated values of Pugh's ratio, Poisson's ratio and Cauchy parameter reveal that RbBaX$_3$ halides are ductile in nature at both ambient condition and under elevated pressure.  

The calculated PBE-GGA band gaps of RbBaF$_3$, 
RbBaCl$_3$, RbBaBr$_3$ and RbBaI$_3$ are 4.8, 4.4, 3.7 and 3.2 eV, respectively.
The spin-orbit coupling effect is observed to be small for the fluoride perovskite, but for other halide perovskites, SOC introduces a 22 meV $\sim$ 311 meV shift in the valance band.
With hybrid HSE correction, the band gap energy increases to 6.7, 5.6, 4.8 eV and 4.4 eV for the F, Cl, Br, and I, respectively.
Among them, RbBaF$_3$ is the only direct band gap material.
However, with external hydrostatic pressure, RbBaCl$_3$ and RbBaI$_3$ become direct band gap perovskites because of a pressure-induced band transition.
Furthermore, the band gap reduces with increasing compressive pressure, showing  tunable band modification for these halides.
The calculated density of states shows that the valance band is controlled by halogen \textit{p}-orbitals, and thus valance bands are more susceptible to splitting of degenerate bands due to spin-orbit coupling, specifically for the larger halogens.

The complex dielectric response and the optical properties of RbBaX$_3$ perovskites are also evaluated using \textit{ab initio} methods. 
The static dielectric constants for the RbBaX$_3$ structures are in the range of 2.4 to 2.8.
All these materials exhibit strong optical absorption in the UV and deep-UV regions, which shows a red-shift towards the visible spectrum from fluoride to iodide. 
External pressure increases the strength and the span of absorption significantly, while introducing a blue-shift in the absorption peaks.
All the RbBaX$_3$ materials exhibit low electron loss, moderate reflectivity and lower refractive index in the UV regions.
External pressure increases the peak strength of these optical properties, and introduces a blue-shift in the respective spectrum.
Such capability of altering the optical absorption spectra, loss function, reflectivity and refractive index makes RbBaX$_3$ perovskites a very good candidate for tunable UV optoelectronics.

\printcredits

\section*{Declaration of competing interest}
The authors declare that they have no known competing financial interests or personal relationships that could have appeared to influence
the work reported in this paper.


\bibliographystyle{vancouver}

\end{document}